\documentclass[twocolumn,showpacs,preprintnumbers,amsmath,amssymb]{revtex4-1}

\usepackage{hyperref}
\usepackage{amsmath}
\usepackage{graphicx}
\usepackage{dcolumn}
\usepackage{simplewick}
\usepackage{color}
\usepackage{bm}

\def\gdphi{\langle\Phi_{v_\gamma w_\delta}|}

\def\phigd{|\Phi_{v_\gamma w_\delta}\rangle}

\begin{document}

\title{Fock space relativistic coupled-Cluster calculations of Two-Valence 
       Atoms}
\author{B. K. Mani and D. Angom}                                                
\affiliation{Physical Research Laboratory,
             Navarangpura-380009, Gujarat,
             India}

\begin{abstract}
   We have developed an all particle Fock-space relativistic coupled-cluster
method for two-valence atomic systems. We then describe a scheme to employ the 
coupled-cluster wave function to calculate atomic properties. Based on these
developments we calculate the excitation energies, magnetic hyperfine structure
constants and electric dipole matrix elements of Sr, Ba and Yb. Further more,
we calculate the electric quadrupole HFS constants  and the
electric dipole matrix elements of Sr$^+$, Ba$^+$ and Yb$^+$. For these we 
use the one-valence coupled-cluster wave functions obtained as an 
intermediate in the two-valence calculations. We also calculate the magnetic
dipole hyperfine structure constants of Yb$^+$.
\end{abstract}



\maketitle

\section{Introduction}

  Coupled-cluster theory, first developed in nuclear many body physics
\cite{Coester-58,Coester-60}, is considered one of the best many body theory. 
In recent times, it has been used with great success in nuclear 
\cite{Hagen-08}, atomic \cite{Nataraj-08,Rupsi-07}, molecular \cite{Isaev-04} 
and condensed matter \cite{Bishop-09} calculations. A recent review 
\cite{Bartlett-07} provides a detailed overview of the theory and 
variations suitable to different classes of many-body systems. An earlier
review provides an overview on the application of coupled-cluster theory to
various areas of physics \cite{Bishop-91}. In atoms it is equivalent 
to incorporating electron correlation effects to all order. It has been used 
extensively in precision atomic structure and properties calculations. These 
include atomic electric dipole moments \cite{Nataraj-08,Latha-09}, parity 
nonconservation \cite{Wansbeek-08}, hyperfine structure constants 
\cite{Rupsi-07,Sahoo-09} and electromagnetic transition properties 
\cite{Thierfelder-09,Sahoo-09a}.

 In this paper we report the development and results of relativistic 
coupled-cluster atomic calculations for two-valence atoms. For this 
we employ the Fock-space open-shell CCT \cite{Mukherjee-79,Lindgren-85}, 
which is also referred as valence universal. Based on which the two-valence CC 
wave operators are calculated via the closed and one-valence wave operators. 
The necessary developments of these intermediate stages of calculations were 
reported in our previous works \cite{Mani-09,Mani-10}. We emphasize that in 
ref. \cite{Mani-10} we proposed a new scheme to calculate properties with CC 
wave functions to all order. In the present work we implement two-valence 
Fock-space relativistic CCT with an all particle valence space. A similar 
approach was adopted in a previous work on relativistic coupled-cluster 
calculations of two-valence systems \cite{Eliav-95}. This is general enough for
the precise wave function and properties calculations of the low-lying levels 
of two-valence systems like the alkaline-earth metal atoms, Yb and Hg. In these
systems, the low-lying levels arise from the $ns^2$, $ns(n-1)d$ and $nsnp$ 
configurations. We show selecting a model space consisting of these 
configurations is incomplete but quasi-complete. Advantage of quasi-complete 
model space is, it has all the virtues of a complete model space but one can 
circumvent the divergence associated with {\em intruder} states \cite{Hose-79} 
in open-shell CCT.  The calculations presented in this work are based on  
coupled-cluster singles and doubles (CCSD) approximation. It was initially 
formulated for molecular calculations \cite{Purvis-82} and used in atomic 
structure calculations to study the excitation energies of Li 
\cite{Lindgren-85pra}. Later, the relativistic version was implemented to 
calculate structure and properties of high $Z$ atoms and ions 
\cite{Ynnerman-94,Ephraim-94,Rupsi-07}. 

 In the present work we apply the method we have developed to calculate
the wave functions of alkaline-Earth atoms Sr and Ba, and lanthanide atom
Yb. All of these atoms are candidates of extremely precise experiments either
for application oriented investigations or to probe fundamental laws of nature.
Atomic Sr, which was recently cooled to quantum degeneracy \cite{Simon-09},
is a strong contender of future optical clocks \cite{Tomoya-08,Ludlow-08}. 
Experiments on Bose-Einstein statistics violations have used Ba as
the target atom \cite{English-10} and it is an ideal proxy, both for 
experimental \cite{De-10} and theoretical \cite{Dzuba-06} studies, of atomic 
Ra. An atom with large parity and time reversal violation effects 
\cite{Flambaum-99}, and promising candidate for future experiments. Recently, 
parity violation was detected in Yb \cite{Tsigutkin-09} and ongoing experiments
could lead to unambiguous detection of nuclear anapole \cite{Zeldovich-58}.
An exotic parity violating nuclear moment, which can be detected only through 
atomic experiments. There are also proposals to measure atomic electric
dipole moment, a signature of parity and time violations, with novel
techniques \cite{Takahashi-97,Natarajan-05}. It must be mentioned that
several isotopes of Yb has been cooled to degeneracy \cite{Takasu-03}
and could be employed in future precision measurements. Further more, atomic Yb
in an optical lattice is a candidate of frequency standard \cite{Barber-06}. 

 As mentioned before, to obtain the two-valence wave operator, we compute
the closed-shell and one-valence wave operators in the intermediate steps.
We take advantage of this and use the one-valence wave operator to compute
the hyper fine constants of Sr$^+$, Ba$^+$ and Yb$^+$. For the first two 
ions, we reported the magnetic hyperfine structure constants (HFS) in our 
previous paper \cite{Mani-10}, so we compute only the electric quadrupole 
HFS constants. Whereas for Yb$^+$ we compute both the magnetic dipole and 
electric quadrupole HFS constants. In addition, we also compute the 
electric dipole transitions matrix elements. Like the neutral atoms,
all the ions are under experimental investigations for various
precision measurements. For example, a single trapped $^{87}{\rm Sr}^+ $ is 
a suitable frequency standard \cite{Barwood-03}. There are similar experiments
with Yb$^+$ \cite{Stenger-02} as an alternative frequency standard. And it is 
one of the frequency standards in the laboratory measurement of temporal
variation of fine structure constant \cite{Peik-04}. These are application 
oriented precision experiments. The other fascinating prospect is the 
observation of parity nonconservation in a single $^{137}{\rm Ba}^+$ 
\cite{Fortson-93}. 

  The paper is divided into seven sections. In the next section, that is 
Section.II,  we give a brief description of many-body perturbation theory
(MBPT) for two-valence  systems. It provides the minimal description of 
concepts pertinent to development of two-valence coupled-cluster theory.
Section.III is a short writeup on closed-shell and one-valence CCT followed
by derivation of the two-valence CCT in some detail. These are in the 
context of complete model space. Incomplete model space CCT is explained
in Section.IV and atomic Yb is discussed as an example. Calculation of 
properties, HFS constants and electric dipole transition,
with CC wave functions is the topic of Section.V. In Section.VI, the important
details of implementing two-valence CCT is explained, however, with emphasis
on physics. Finally, results and discussions are reported in Section.VII. In 
the paper, all the calculations and mathematical expressions are in atomic
units ($e=\hbar=m_e=4\pi\epsilon_0=1$).


\section{MBPT for two-valence atoms}

  Relativistic effects are the key to obtain accurate results in the structure 
and properties calculations of high $Z$ atoms with $z\alpha\sim 1 $. The 
Dirac-Coulomb Hamiltonian $H^{\rm DC}$ is an approximate but an  appropriate 
Hamiltonian to describe the properties of such atoms. For an atom with $N$ 
electrons
\begin{equation}
  H^{\rm DC}=\sum_{i=1}^N\left [c\bm{\alpha}_i\cdot \bm{p}_i+
             (\beta-1)c^2 - V_N(r_i)\right ] +\sum_{i<j}\frac{1}{r_{ij}},
  \label{dchamil}
\end{equation}
where $\bm{\alpha}_i$ and $\beta$ are the Dirac matrices, $\bm p$ is the 
linear momentum, $V_N(r)$ is the nuclear Coulomb potential and last term
is the electron-electron Coulomb interactions. It satisfies, in the case of 
two-valence atoms, the eigen value equation
\begin{equation}
  H^{\rm DC}|\Psi_{vw}\rangle = E_{vw}|\Psi_{vw}\rangle,
  \label{dcpsi}
\end{equation}
where indexes $v$ and $w$ represent the valence orbitals, $|\Psi_{vw}\rangle$ 
is the exact wave function and $E_{vw}$ is the exact energy of the 
two-valence atomic system. In MBPT, the total Hamiltonian, in 
Eq.(\ref{dchamil}), is separated into two parts: 
$H_0 = \sum_i[c\bm{\alpha}_i\cdot\bm{p}_i + (\beta_i-1)c^2-V_N{r_i} + 
u(\bm{r}_i)]$, the unperturbed or exactly solvable part, and 
$V=\sum_{i<j}^N\frac{1}{\bm{r}_{ij}}-\sum_i u(\bm{r}_i)$, the perturbation,
referred as the residual Coulomb interaction. The unperturbed 
eigen functions $|\Phi_{vw}\rangle$ are the solution of the Dirac-Fock equation
\begin{equation}
  H_0|\Phi_{vw}\rangle = E^{(0)}_{vw} |\Phi_{vw}\rangle , 
\end{equation}
here, $|\Phi_{vw}\rangle$ are the antysymmetrised many-electron wave functions. 
Formally, in operator notations, these are generated from the closed-shell 
reference state $|\Phi_0\rangle $ as  
$|\Phi_{vw}\rangle=a^{\dagger}_va^{\dagger}|\Phi_0\rangle$.
And, the eigen value, $E^{(0)}_{vw}$, is the sum of the single electron 
energies. These are the basic starting points common to 
atomic MBPT and coupled-cluster theory (CCT). The two theories share a common 
thread till the generalized Bloch equation \cite{Lindgren-74}, discussed in 
the next section, but is significantly different from there on.


\subsection{Generalized Bloch equation}

 In this section, and others as well, we provide the basic 
equations and necessary definitions essential to a lucid description of the 
method we have developed and used. For detailed descriptions appropriate 
references are provided. In MBPT, the Hilbert space of the eigen functions 
$|\Phi_{vw}\rangle$, is separated into two sub-manifolds: model space ($P$), 
comprise of the eigen functions which are the best approximation to the exact 
eigen functions of interest and remaining spans the orthogonal space ($Q$). 
In the single reference theory, the exact state $|\Psi_{vw}\rangle$ and model 
state $|\Phi_{vw}\rangle$ are related as
\begin{equation}
  |\Psi_{vw}\rangle = \Omega |\Phi_{vw}\rangle,   
  \label{exact_state} 
\end{equation}
where, $ \Omega$ is the wave operator and is the solution of the generalized
Bloch equation
\begin{equation}
  [\Omega,H_0]P = (V\Omega - \Omega PV\Omega)P.
  \label{bloch_eq}
\end{equation}
Detailed exposition of the equation and relevant derivations are given 
ref. \cite{Lindgren-74,Lindgren-85ps}. Here, intermediate normalization
\begin{equation}
  |\Phi_{vw}\rangle =  P\Omega|\Phi_{vw}\rangle,  
 \label{int_norm} 
\end{equation}
is a necessary condition to obtain the generalized Bloch equation.
In singles and doubles approximation, 
often used and well tested method in atomic calculations, the wave operator is
\begin{eqnarray}
  \Omega &=& 1 + x_a^p a^{\dagger}_pa_a + x_v^p a^{\dagger}_pa_v 
             + \frac{1}{2}x_{ab}^{pq}a_p^{\dagger}a_q^{\dagger} a_ba_a
             + x_{av}^{pq}a_p^{\dagger}a_q^{\dagger} a_va_a \nonumber \\
         & & + \frac{1}{2}x_{vw}^{pq}a_p^{\dagger}a_q^{\dagger} a_wa_v,
  \label{exc_op}
\end{eqnarray}
where $ab\cdots (pq\cdots)$ denote core (virtual) orbitals and 
$x_{\cdots}^{\cdots}$ are the excitation amplitudes. This definition is crucial
to our later discussions on the Fock space coupled-cluster in complete model 
space (CMS). Unlike the close-shell atoms, the model 
wave functions are not known in the case of open-shell systems. These are 
obtained by diagonalizing the effective Hamiltonian ($H_{\rm eff}$ ) matrix, 
calculated within the $P$ sub-manifold. Once the model wave function is 
obtained, the exact energy is the expectation value of $H_{\rm eff}$, as it 
satisfies the eigen value equation
\begin{equation}
  H_{\rm eff}|\Phi_{vw}\rangle = E_{vw} |\Phi_{vw}\rangle.
   \label{h_eff1}
\end{equation}
The effective Hamiltonian, in Eq.(\ref{h_eff1}), is expressed as
\begin{equation}
   H_{\rm eff} = PH_0P + PV\Omega P,
   \label{h_eff2}
\end{equation}
where $H_0$ and $V$, as defined earlier, are the zeroth-order Hamiltonian and 
the residual Coulomb interaction respectively. The first term in 
Eq.(\ref{h_eff2}) is the leading contribution, $E^{(0)}_{vw}$, to the total 
energy $E_{vw}$. And the second term, with the wave operator $\Omega$, 
is the correction to $E^{(0)}_{vw}$ referred as correlation energy.


\subsection{First- and second-order effective Hamiltonians}

\begin{figure}[h]
\begin{center}
  \includegraphics[width = 8.2cm]{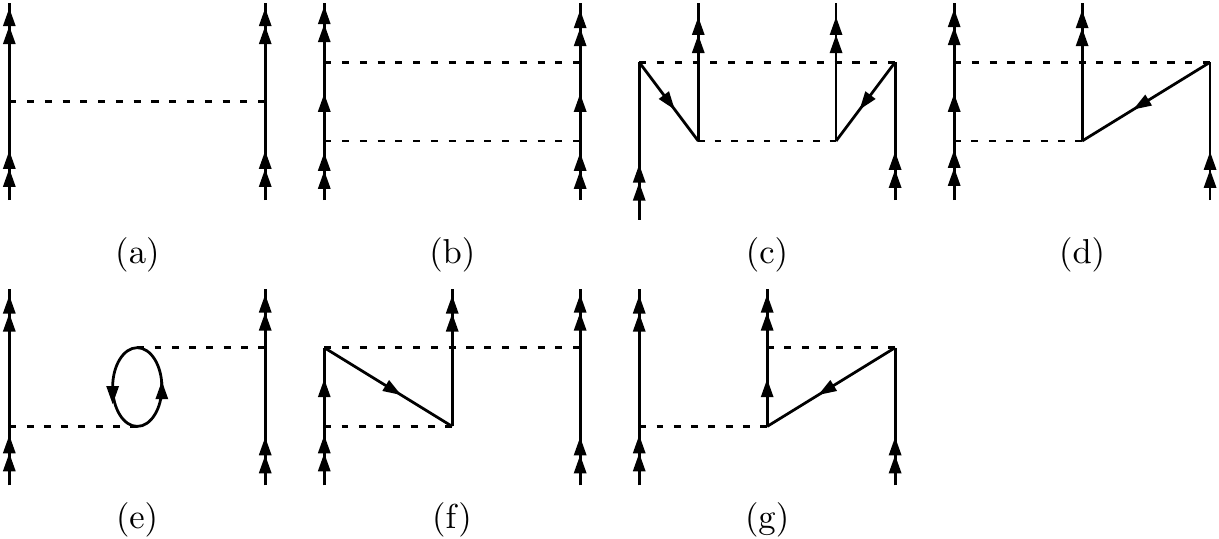}
  \caption{The two-body diagram $(a)$, arises from the first-order effective 
           Hamiltonian $H^{(1)}_{\rm eff}$. The remaining two-body diagrams,
           from $(b) - (g)$, contribute to the second-order effective
           Hamiltonian $H^{(2)}_{\rm eff}$.}
  \label{doubles_2v}
\end{center}
\end{figure}

 From Eq.(\ref{h_eff1}), the first-order correction to energy is the 
expectation value of the first-order effective Hamiltonian 
\begin{equation}
  H^{(1)}_{\rm eff} = PVP = P(V_0 + V_1 + V_2)P.
  \label{h_eff3}
\end{equation}
where, $V_0$ is the contribution from the close-shell part and represented 
by the closed diagrams with no free lines at the vertexes. We exclude this 
term while calculating the excitation energies as it is common 
to all the diagonal elements of the $H_{\rm eff}$ matrix. It 
effectively shifts all the energy levels equally and does not account for the 
energy level splitting. The one- and two-body terms, $V_1$ and $V_2$, have 
contributions from open-shell part only. The contributing diagrams are the 
closed diagrams with free valence lines at the vertixes. The one-body term, 
$V_1$, also contributes to the diagonal elements only and hence does not 
contribute to the energy level splitting.  From Eq.(\ref{h_eff3}), 
$H^{(1)}_{\rm eff}$ is reduced to the form
\begin{equation}
  H^{(1)}_{\rm eff} = PV_2P.
  \label{h_eff1b}
\end{equation}
This term contributes through a closed diagram with one pair of valence
lines at each vertex shown in Fig. \ref{doubles_2v}(a).
\begin{figure}[h]
\begin{center}
  \includegraphics[width = 8.0cm]{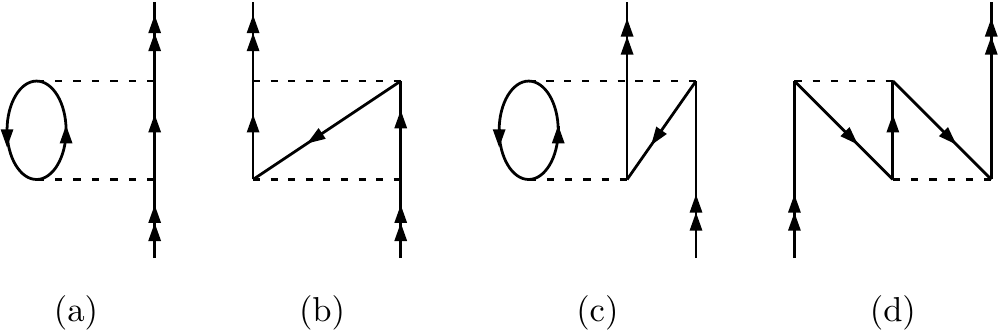}
  \caption{One-body diagrams arising from the second-order effective
           Hamiltonian $H^{(2)}_{\rm eff}$.}
  \label{singles_2v}
\end{center}
\end{figure}
  From Eq.(\ref{h_eff2}), the second-order effective Hamiltonian
\begin{equation}
  H^{(2)}_{\rm eff} = PV\Omega^{(1)}P = P(V_1+V_2)(\Omega^{(1)}_1+
  \Omega^{(1)}_2)P, 
  \label{h_eff4}
\end{equation}
where the superscript in the wave operator represents the order of the
perturbation. Contributing diagrams are closed diagrams with valence orbitals
as free lines at the vertexes. Detailed description of the relativistic
second MBPT of two-valence systems is given ref. \cite{Safronova-96}.
The terms involving $V_1$ are
zero if Dirac-Fock orbitals are used in the calculations. The expression of
$H^{(2)}_{\rm eff}$ is 
\begin{equation}
  H^{(2)}_{\rm eff} = P \contraction[0.5ex]{}{V}{_2}{\Omega}
                         V_2\Omega^{(1)}_2 P. 
  \label{h_eff2b}
\end{equation}
where $\{\contraction[0.5ex]{}{A}{\cdots}{B}A\cdots B\}$  represents 
contraction between two operators $A$ and $B$. 
The diagrams of $H^{(2)}_{\rm eff}$, in Eq.(\ref{h_eff2b}), are separated
into two categories. The diagrams with one pair of free lines as the valence
orbitals, shown in Fig. \ref{singles_2v}, constitute the one-body effective
operator. And the diagrams with two pair of free lines as the valence
orbitals, shown in Fig. \ref{doubles_2v}, forms the two-body effective 
operator. It must be mentioned that, a previous work reported the third 
ordered relativistic MBPT calculations of two-valence systems beryllium
and magnesium iso-electronic sequences \cite{Ho-06}.


\subsection{$H_{\rm eff}$ matrix elements with $jj$ coupled states} 

  In our scheme of calculations, we first evaluate the diagrams arising  from
Eqs.(\ref{h_eff1b}) and (\ref{h_eff2b}), Figs. \ref{singles_2v} and 
(\ref{doubles_2v}), using uncoupled states. And we then store these as the 
effective one- and two-body operators. Later we use these effective operators 
to generate the matrix elements with respect to the $jj$ coupled states. For 
two non-equivalent electrons the $jj$ coupled antysymmetrised state may be 
expressed, in terms of the total angular momentum $J$ state, as
\begin{align} 
  &|\{\gamma_vj_vm_v\gamma_wj_wm_w\}JM\rangle =
   \frac{1}{\sqrt 2}
  \Bigl[|(\gamma_vj_vm_v\gamma_wj_wm_w)JM\rangle \Bigr. \nonumber \\
  & \;\;\;\;\; + (-1)^{j_v+j_w+J}
  |(\gamma_wj_wm_w\gamma_vj_vm_v)JM\rangle \Bigr].
   \label{coupled_s}
\end{align} 
where, $j_v$ and $j_w$ are the total angular momenta of the single electron
states $|\phi_{v}\rangle$ and $|\phi_{w}\rangle$ respectively, $\gamma_v$ and 
$\gamma_w$ are additional quantum numbers to specify the states uniquely. 
And $m_v$ and $m_w$ are the corresponding magnetic quantum numbers.
Similarly, $J$ and $M$ are the total angular momentum of the coupled state 
and magnetic quantum number respectively. The matrix element of a 
two-body operator then consists of four terms, two direct and two exchange, 
with the normalization factor $1/2$. 

  To evaluate the two-body matrix element, for example, the coulomb interaction
shown Fig. \ref{doubles_2v}a. The direct matrix element is of the form
\begin{eqnarray}
  &&\langle(\gamma_vj_vm_v \gamma_wj_wm_w)JM|\frac{1}{r_{12}}
  |(\gamma_xj_xm_x \gamma_yj_ym_y)J'M'\rangle
   = \nonumber \\
  &&\sum_k(-1)^{j_x+j_w+J+k}\delta(J,J')\delta(M,M')
    \left\{\begin{array}{ccc}
                     j_x & j_x & k\\
                     j_y& j_w & J
                 \end{array}\right\} \nonumber  \\
  &&\times \langle \gamma_vj_v||\mathbf{C}^k||\gamma_xj_x\rangle 
    \langle \gamma_wj_w||\mathbf{C}^k||\gamma_yj_y\rangle \times R^k.
    \label{cmatrix}
\end{eqnarray}
Where $x$ and $y$ represent valence orbitals, $R^k$ is the radial integral and 
${\bm C}^k$ is the spherical tensor operator. For matrix elements, in 
Eq. (\ref{cmatrix}), to be non-zero the states should have the same parity and 
$J$.  The relation in Eq.(\ref{cmatrix}) holds true for the matrix elements of 
the other two-body diagrams Fig. \ref{doubles_2v}(b-g). In this case the 
multipole $k$ and the radial integral arise from the combination of two orders 
of residual Coulomb interactions.

  Similarly, the matrix element of the one-body operator of rank $k$,
with respect to the $jj$ coupled state is
\begin{eqnarray}
  &&\langle(\gamma_vj_v \gamma_wj_w)JM|\mathbf{F}^k(1)
  |(\gamma_xj_x \gamma_yj_y)J'M'\rangle = \nonumber \\
  &&\delta(\gamma_w,\gamma_y)\delta(j_w,j_y)
    (-1)^{J-M}(-1)^{j_v + j_y + J'+k}[J,J']^{1/2}  \nonumber \\
  &&\left(\begin{array}{ccc}
                   J & k & J'\\
                  -M & 0 & M
                 \end{array}\right) 
    {\left\lbrace\begin{array}{ccc}
                   j_v  & j_x & k  \\
                   J'   & J    & j_w
                 \end{array}\right\rbrace} 
  \langle \gamma_vj_v||\mathbf{f}^k||\gamma_xj_x\rangle.
\end{eqnarray}
This is a very general expression and applicable to one-body operator
of any rank $k$. In our calculations, however, we use $k = 0$ as the
one-body effective operator is scalar.


\section{Fock space CCT: complete model space} 

  A model space is complete, if it consists of all the configurations formed by
accommodating the valence electrons among the valence shells in all possible 
combinations. A remarkable consequence of choosing CMS in Fock-space 
coupled-cluster is that, the excitation operators
$x_{\cdots}^{\cdots}$ are common to all the determinants in the model space. 
Further more, $x_{\cdots}^{\cdots}$ uniquely separates into internal and 
external sectors. The external excitations contribute to $\Omega $ and projects
a model function to the complementary space. Whereas, internal excitations 
connect one model function to another model function and occurs in the 
definition of $H_{\rm eff}$. As we shall explore later, in the 
context of incomplete model space (IMS), such a neat separation is specific to
CMS and another class of model space referred as quasi complete 
\cite{Lindgren-85ps}. Validity of linked cluster theorem, one basic condition
for any legitimate many-body theory, is assured in CMS.

\begin{figure}[h]
\begin{center} 
  \includegraphics[width = 8.2cm]{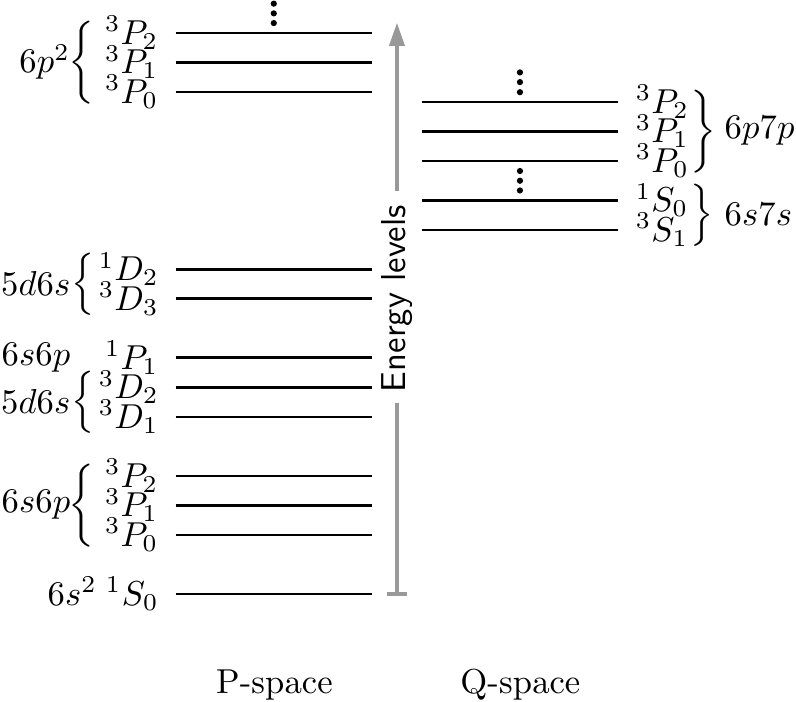} 
  \caption{Low-lying energy levels of atomic Yb.}
  \label{yb_elevel}
\end{center}
\end{figure}

 The CMS, though endowed with several desirable properties, has one 
serious short coming for systems with two or more valence electrons. It 
inevitably encounters {\em intruder} states \cite{Hose-79}  and the outcome 
is severe convergence problems. This is the manifestation of model states
with high energies that lies within the energy domain of the orthogonal space.
In other words, in CMS when all possible configurations are considered,
the model and complementary space are no longer energetically well separated.
The occurrence of vanishing energy denominators is then a distinct
possibility. Indeed, we invariably encounter {\em intruder} states in all our 
two-valence calculations with CMS. Its presence is the rule rather than the 
exception. 

 For a better description of the CMS and {\rm intruder} states let us consider 
a specific example, the low-lying levels of Yb atom. Configurations and terms 
of the ground and first few excited states important in precision spectroscopy 
are $6s^2~(^1S_0)$, $6s6p~(^3P_J)$, $5d6s~(^3D_J)$ and $6s6p~(^1P_1)$. The 
$6s$, $6p$ and $5d$ are then the obvious choice of valence shells. CMS of the 
system then consists of the configurations: $6s^2$, $6s6p$, $5d6s$, $5d6p$, 
$6p^2$ and $5d^2$ and all the other configurations are in the complementary
space. As shown in Fig. \ref{yb_elevel}, the levels from the orthogonal space 
$6p7p~(^3P_J)$, $6s7s~(^3S_1)$ and $6s7s~(^1S_0)$ lie within  the model space. 
With several orthogonal functions within the energy domain of model functions, 
CMS based CCT calculations of Yb are likely to face with {\rm intruder} state 
related divergences. Indeed, we do encounter divergences which, on careful 
analysis, can be attributed to the {\rm intruder} states.

  In this work, we proceed to the relativistic two-valence 
coupled-cluster theory via the closed-shell \cite{Mani-09} and one-valence 
coupled-cluster \cite{Mani-10} theories reported in our previous works. We 
implement this within the framework of Fock-space or valence universal CCT
\cite{Mukherjee-75,Mukhopadhyay-79}. The theory can be extended to systems with
both particles and holes, however, for our present study an all particle 
implementation is sufficient.  Accordingly, the valence electrons are treated 
as particles \cite{Lindgren-78} and each sector--closed-shell, one-
and two-valence--are separate Hilbert spaces. Technical advantage of 
Fock-space CCT with CMS is the sector wise clean separation of cluster 
operators \cite{Mukherjee-86}. However, the Hilbert spaces of two-valence 
subsumes the one-valence after a direct product with a spectator valence 
state and similarly, the closed-shell after direct product with two-valence 
states. The state universal \cite{Jeziorski-81} is another flavour of 
open-shell CCT, where the wave operator is calculates in a single Hilbert 
space consisting of all the valence electrons. The wave operator is then
state dependent and there is a lack of generality in the cluster equations.
Further more, it requires complicated book keeping.  These reasons have 
motivated us to choose the Fock-space CCT. Detailed discussions on Fock-space
CCT and subtle issues related to the choice of model spaces are given in the 
review of Lindgren and Mukherjee \cite{Lindgren-87}.

\begin{figure}[h]
\begin{center} 
  \includegraphics[width = 8.2cm]{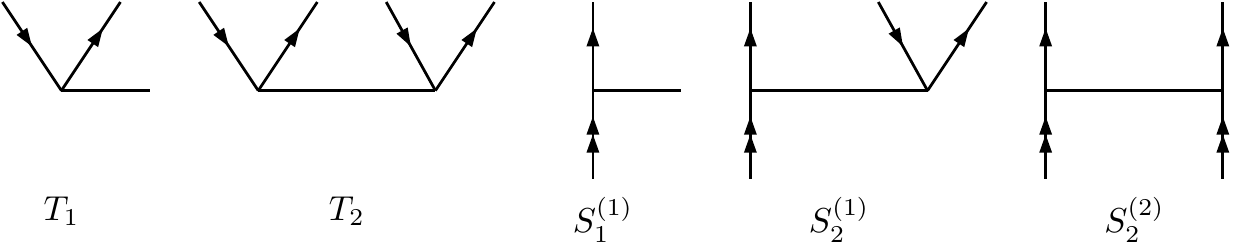} 
  \caption{Representation of the closed-shell and open-shell cluster operators}
  \label{coperators}
\end{center}
\end{figure}


\subsection{Closed-shell and one-valence CCT }

   Coupled-cluster theory is a non-perturbative many-body theory.
It is equivalent to selecting linked terms in the Bloch equation, 
Eq.(\ref{bloch_eq}), to all orders and combining terms of same level of 
excitation (LOE). 
The elegance and perhaps, all the attendant difficulties of CCT is the 
exponential nature of the wave operator. We 
provide a brief reprise of the closed-shell \cite{Mani-09}
and one-valence CCT \cite{Mani-10}, these form the initial steps of the 
two-valence coupled-cluster theory. Indeed, a large fraction of the cluster 
amplitudes in the two-valence theory arise from the close-shell.
The exact atomic wave function of a one-valence system
in the coupled-cluster theory is
\begin{equation}
  |\Psi_{v}\rangle = e^{(T + S)} = e^T(1+S)|\Phi_{v}\rangle ,
  \label{onev_exact}
\end{equation}
where $|\Phi_{v}\rangle$ is the one-valence  Dirac-Fock reference state,
and $T$ and $S$ are the closed- and open-shell cluster operators respectively. 
As evident from the equation, all the higher order terms (non-linear) of $T$ 
are calculated but $S$ restricted to linear terms only. The later is on account
of the single valence electron. The diagrammatic representations of 
these operators are shown in Fig. \ref{coperators}.

  The closed-shell operator $T$ in the coupled-cluster singles doubles (CCSD) 
\cite{Purvis-82} approximation is 
\begin{equation}
  T = T_1 + T_2,
\end{equation}
where $T_1$ and $T_2$ are the single and double excitation operator
respectively. The closed-shell exact state in CCT is 
\begin{equation}
  |\Psi_0\rangle = e^T|\Phi_0\rangle,
\end{equation}
and the cluster amplitudes are solutions of the nonlinear coupled equations
\begin{eqnarray}
  \langle\Phi^p_a|\bar H_{\rm N}|\Phi_0\rangle = 0, 
     \label{t1_eqn}                        \\
  \langle\Phi^{pq}_{ab}|\bar H_{\rm N}|\Phi_0\rangle = 0,
     \label{t2_eqn} 
\end{eqnarray}
where $\bar H_{\rm N}=e^{-T}H_{\rm N}e^{T} $ is the
similarity transformed or dressed Hamiltonian. $|\Phi_0\rangle$, the 
Dirac-Fock reference state for the closed shell part, is the eigen value of
the central potential Hamiltonian $H_0$. And $|\Phi^p_a\rangle$ and 
$|\Phi^{pq}_{ab}\rangle$ are respectively, the singly and doubly excited 
determinants. For details of the derivation, readers are referred to ref.
\cite{Mani-09}. The open-shell cluster operator $S$ is 
\begin{equation}
  S = S^{(1)} + S^{(2)},
  \label{s_op}
\end{equation}
where $S^{(1)}$ and  $S^{(2)}$  are the one-valence and two-valence 
cluster operators respectively. Similar to $T$, the open-shell one-valence 
cluster operator $S^{(1)}$, in CCSD approximation, is of the form 
$S^{(1)} = S^{(1)}_1 + S^{(1)}_2$. And these are solutions of the 
coupled linear equations
\begin{eqnarray}
  \langle \Phi_v^p|\bar H_N \! +\! \{\contraction[0.5ex]
  {\bar}{H}{_N}{S} \bar H_N S^{(1)}\} |\Phi_v\rangle
  &=&E_v^{\rm att}\langle\Phi_v^p|S^{(1)}_1|\Phi_v\rangle ,
  \label{ccsingles}     \\
  \langle \Phi_{va}^{pq}|\bar H_N +\{\contraction[0.5ex]
  {\bar}{H}{_N}{S}\bar H_N S^{(1)}\} |\Phi_v\rangle
  &=& E_v^{\rm att}\langle\Phi_{va}^{pq}|S^{(1)}_2|\Phi_v\rangle.
  \label{ccdoubles}
\end{eqnarray}
In these equations $E_v^{\rm att}$ is the attachment energy of an electron
to the $v$ shell. It is defined as 
\begin{equation}
   E_v^{\rm att} = E_v - E_0,
\end{equation}
where $E_v = \langle \Phi_v|\bar H_N + \{\contraction[0.5ex]
  {\bar}{H}{_N}{S} \bar H_N S^{(1)}\} |\Phi_v\rangle$  and
$E_0=\langle\Phi_0|\bar H |\Phi_0\rangle$  these are the exact energy of 
$|\Psi_v\rangle$ and $|\Psi_0\rangle$ respectively. The excited determinants, 
$|\Phi^p_v\rangle$ and $|\Phi^{pq}_{va}\rangle$, are obtained by exciting an 
electron from valence orbitals to the virtuals. For detail description, of 
the derivation and interpretations, one may see ref \cite{Mani-10}.
\begin{figure}[h]
\begin{center}
  \includegraphics[width = 8.2cm]{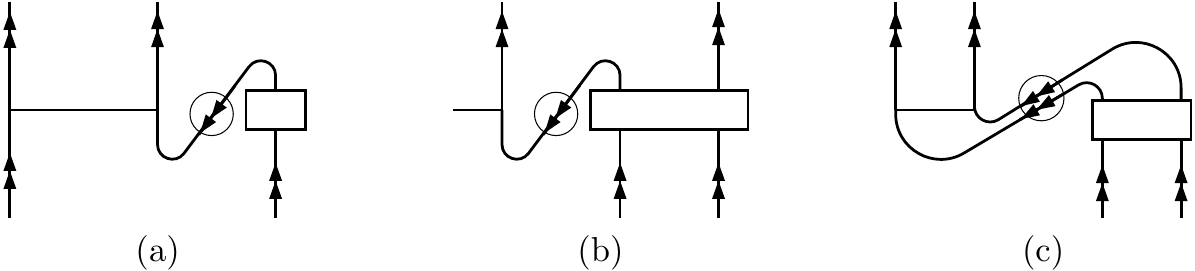}
  \caption{Folded diagrams from the renormalization term in the generalized
           Bloch equation of two-valence systems. In two-valence 
           coupled-cluster theory these diagrams arise from (a) 
           $E_{vw}^{\rm att}\langle\Phi_{vw}^{pq}|S_2^{(1)}|\Psi_{vw}\rangle$,
           (b)$E_{vw}^{\rm att}\langle\Phi_{vw}^{pq}|S_1^{(1)}|\Psi_{vw}\rangle$
           and (c)$E_{vw}^{\rm att}\langle\Phi_{vw}^{pq}|S_2^{(2)}|\Psi_{vw}
           \rangle$.
           }
  \label{folded_diag}
\end{center}
\end{figure}
Nonzero renormalization, right hand side in 
Eq. (\ref{ccsingles}-\ref{ccdoubles}), is the predominant departure  of 
open-shell CC from closed-shell CC. In the language of many-body diagrams,
folded diagrams embody the renormalization terms. These, the folded diagrams,
are topologically very different from the diagrams of $\bar H$ or 
$\contraction[0.5ex] {\bar}{H}{_N}{S}\bar H_N S^{(1)}$. To illustrate the
difference folded diagrams from the two-valence CC are shown in 
Fig. \ref{folded_diag}. Strictly speaking, the distortion in these diagrams
are introduced to obtain correct energy denominators with the diagrammatic
evaluation in MBPT. CC being non-perturbative there is no reason to be 
concerned about correct denominators. However, we retain the nomenclature and 
structure, in the diagrammatic analysis of CC equations, to identify the 
diagrams uniquely.


\subsection{Two-valence CCT}

   In the two-valence sector, the cluster operator $S^{(2)} = S^{(2)}_2$, a 
natural outcome of treating the single valence excitations as one-valence 
problem in Fock-space CCT. The exact two-valence state in CCT is
\begin{equation}
   |\Psi_{vw}\rangle = e^T \left[ 1 + S^{(1)}_1 + \frac{1}{2}{S^{(1)}_1}^2 + 
                       S^{(1)}_2 + S^{(2)}_2\right ]|\Phi_{vw}\rangle .
  \label{twov_exact}
\end{equation}
Here, for the valence part we have used $ \exp(S)= 1 + S^{(1)}_1 + 
(1/2){S^{(1)}_1}^2 + S^{(1)}_2 + S^{(2)}_2$. Notice that, though 
$(1/2){S^{(1)}_1}^2$  does not contribute to the one-valence CC equations, it 
does contribute to the two-valence CC equations. Using this in Eq.(\ref{dcpsi})
and projecting on $e^{-T}$, we get
\begin{eqnarray}
  && \bar H \left [ 1 + S^{(1)}_1 + \frac{1}{2}{S^{(1)}_1}^2 + 
      S^{(1)}_2 + S^{(2)}_2 \right ]|\Phi_{vw}\rangle 
  = E_{vw}\biggl [1  \biggr . \nonumber \\
  && \biggl. + S^{(1)}_1 + \frac{1}{2}{S^{(1)}_1}^2 + 
      S^{(1)}_2 + S^{(2)}_2   \biggr ]|\Phi_{vw}\rangle.
\end{eqnarray}
Here after for simplicity of representation we use 
$\exp(S) = 1 + S + (1/2)S^2$ with the definition, restricted to two-valence
sector only, $S^2 = {S_1^{(1)}}^2$. Using the normal-ordered form the 
Hamiltonian, $H = H_{\rm N}+E^{\rm DF}_{vw}$, we can write
\begin{eqnarray}
   && \bar H_{\rm N}\left [ 1 + S + \frac{1}{2}S^2 \right ] 
      |\Phi_{vw}\rangle = \Delta E^{\rm corr}_{vw}\biggl [ 1 + S \biggr .
            \nonumber \\
   && \biggl. +\frac{1}{2}S^2 \biggr ]|\Phi_{vw}\rangle,
  \label{hnormal}
\end{eqnarray}
where $\Delta E^{\rm corr}_{vw} = E_{vw} - E^{\rm DF}_{vw},$ is 
the correlation energy of the two-valence atoms and as defined earlier,
Eq.(\ref{s_op}), $S=S^{(1)} + S^{(2)}$. Projecting above 
equation with $\langle\Phi_{vw}|$, we get the following expression for the
correlation energy
\begin{equation}
  \langle\Phi_{vw}|\bar H_{\rm N}\left [ 1 + S + \frac{1}{2}S^2
    \right ] |\Phi_{vw}\rangle = \Delta E^{\rm corr}_{vw}.
  \label{ecorr}
\end{equation}
On the right hand side we have used the relations 
$\langle\Phi_{vw}|S|\Phi_{vw}\rangle = 0$ and
 $\langle\Phi_{vw}|S^2|\Phi_{vw}\rangle = 0$, as the operation of $S$ on the
state $|\Phi_{vw}\rangle$ transforms it to an excited determinant
orthogonal to $\langle\Phi_{vw}|$.

 To obtain the two-valence cluster equations, we project Eq.({\ref{hnormal}}) 
on the doubly excited determinants
\begin{eqnarray}
  \langle\Phi^{pq}_{vw}|\bar H_{\rm N}\left [ 1 + S + 
  \frac{1}{2}S^2\right ] |\Phi_{vw}\rangle =   \nonumber \\
   \Delta E^{\rm corr}_{vw}
   \langle\Phi^{pq}_{vw}|S + \frac{1}{2}S^2|\Phi_{vw}\rangle,
\end{eqnarray}
where we have used $\langle\Phi^{pq}_{vw}|\Phi_{vw}\rangle = 0$. This
equation can further be simplified, using Wick's theorem, as
\begin{eqnarray}
   \langle\Phi^{pq}_{vw}|
    \bar H_{\rm N} +
   \{\contraction{\bar}{H}{_{\rm N}}{S}\bar H_{\rm N}S\}+ 
   \frac{1}{2}\{\contraction{\bar}{H}{_{\rm N}}{S}\bar H_{\rm N}S^2\} 
    |\Phi_{vw}\rangle =      \nonumber \\ 
    E^{\rm att}_{vw}\langle\Phi^{pq}_{vw}|S + \frac{1}{2}S^2
   |\Phi_{vw}\rangle,
   \label{ccsd_2v}
\end{eqnarray}
where $E^{\rm att}_{vw}$ is the difference between the exact energy of the
closed-shell and two-valence states. It has the expression
\begin{equation}
  E^{\rm att}_{vw} = \epsilon_v + \epsilon_w + \Delta E^{\rm att}_{vw},
  \label{2v_eatt}
\end{equation}
where, $\epsilon_v$ and $\epsilon_w$ are the Dirac-Fock energy of the valence
orbitals $|\phi_v\rangle$ and $|\phi_w\rangle$ respectively. And
$\Delta E^{\rm att}_{vw} = \Delta E^{\rm corr}_{vw}-
\Delta E^{\rm corr}_0$, is the difference of the correlation energy of
closed-shell  and two-valence states. Diagrammatically,
$\Delta E^{\rm att}_{vw}$ in Eq.(\ref{2v_eatt}) is equivalent to the closed 
diagrams with free lines representing the valence states at the vertexes. Like 
in the second order MBPT, one can separate the $\Delta E^{\rm att}_{vw}$ 
diagrams to one- and two-body types. The one-body diagrams are similar to
the ones in Fig. \ref{singles_2v} with the bottom interaction (dotted line)
replaced by cluster amplitude (solid line). Similarly, the two-body diagrams
are  similar to those of Fig. \ref{doubles_2v}(b-g) with the bottom
interaction replaced by the cluster amplitude.


\subsection{CC equation from Bloch equation}

  The CC equations discussed so far are derived from the
eigenvalue equation of the Dirac-Coulomb Hamiltonian. Another approach is 
based on the generalized Bloch equation given in Eq. (\ref{bloch_eq}). This is 
more transparent to implement and convenient to analyse the working equations 
of CC with incomplete model space. In Eq. (\ref{bloch_eq}), the second term
on the right hand side, renormalization term, is often defined as
\begin{equation}
  W = P V\Omega P = (V\Omega)_{\rm close}.
  \label{w_eff}
\end{equation}
Here, {\em close } indicates the operator connects states within 
the model space. Diagrammatically, the representation of the operator has no 
free lines in the closed-shell sector and only valence orbitals as free lines 
in the open-shell sector. Using Eq.(\ref{w_eff}), we can write
\begin{equation}
  [\Omega,H_0]P = (V\Omega - \Omega W)P.
  \label{bloch_eq2}
\end{equation}
Operating on the two-valence atomic reference state, $|\Phi_{vw}\rangle$, and 
projecting with the doubly excited determinant $\langle\Phi^{pq}_{vw}|$, we get
\begin{eqnarray}
  && \langle\Phi^{pq}_{vw}|[e^{T + S},H_0]|\Phi_{vw}\rangle = \langle
     \Phi^{pq}_{vw}|\left[Ve^T\left(1 + S + \frac{1}{2}S^2\right)\right .
                  \nonumber \\
  && \left. - e^T\left( 1 + S + \frac{1}{2} S^2\right)W\right]|\Phi_{vw}\rangle.
  \label{bloch_eq2}
\end{eqnarray}
From Wick's theorem further simplification follows after contracting the 
operators. There are connected and disconnected terms, however, only the 
connected terms remain \cite{Lindgren-85} on both sides of 
Eq.(\ref{bloch_eq2}). We get
\begin{eqnarray}  
  && \langle\Phi^{pq}_{vw}| \{\contraction{}{H}{_0}{S}H_0S\} - 
     \{\contraction{}{S}{}{H}SH_0\}|\Phi_{vw}\rangle = -\langle\Phi^{pq}_{vw}
     |\biggl[Ve^T\biggl(1 + S  +   \biggr. \biggr .
                       \nonumber \\ 
  && \left.\left. \frac{1}{2}S^2\right) - e^T\left(1 + S + 
     \frac{1}{2}S^2\right) W\right]_{\rm conn} |\Phi_{vw}\rangle ,
  \label{cc_eq}
\end{eqnarray}  
where the subscript {\em conn} refers to connected terms. To arrive at the 
equation we have used $\langle\Phi^{pq}_{vw}|[T,H_0]|\Phi_{vw}\rangle = 0$, 
as $T$ being the closed-shell cluster operator, it does not operate in the 
valence space. To examine the equation in further detail, consider the terms 
on the right hand side. Expanding the exponential in the term $ Ve^T$
\begin{eqnarray}
  \left ( Ve^T \right )_{\rm conn} & = & V +\{\contraction{}{V}{}{T}VT\} +
  \frac{1}{2!}\{\contraction{}{V}{}{T}
  \contraction[1.5ex]{}{V}{T}{T}VTT\} +
  \frac{1}{3!}\{\contraction{}{V}{}{T}
  \contraction[1.5ex]{}{V}{T}{T}
  \contraction[2.0ex]{}{V}{TT}{T}VTTT\}    \nonumber \\
  && +\frac{1}{4!}\{\contraction{}{V}{}{T}
  \contraction[1.5ex]{}{V}{T}{T}
  \contraction[2.0ex]{}{V}{TT}{T}
  \contraction[2.5ex]{}{V}{TTT}{T}VTTTT\}
  =\bar V,
  \label{cc_eqr1}
\end{eqnarray}
where $\bar V$ is the dressed operator. Similarly, for 
the other terms 
\begin{eqnarray}
  \left ( Ve^TS\right )_{\rm conn} &= &\{\contraction{}{\bar V}{}{S}
               \bar VS\}, \\
  \left( e^TSW \right)_{\rm conn} &= & \{\contraction{}{S}{}{W}SW\}, \\ 
  \left(e^TW\right)_{\rm conn} &= &0,
  \label{cc_eqr2}
\end{eqnarray}
as no contraction can occur between $T$, the closed-shell cluster operator,
and open-shell operator $S$ to obtain connected term. The same is true of
$T$ and the effective interaction $W$. The reason is, $T$ operates on the 
closed-shell sector, whereas $W$ operates in the valence sector. Though it
is not shown explicitly, there are similar relations for $(1/2)S^2 $ as well. 
From the definition of the normal Hamiltonian 
$\bar H_{\rm N} = \bar V + \bar H_0$ we can combine two of the terms as
\begin{equation}
  \contraction{\bar}{H}{_0}{S}\bar H_0S + 
  \contraction{\bar}{V}{}{S}\bar VS = 
  \contraction{\bar}{H}{_{\rm N}}{S}\bar H_{\rm N}S.
\end{equation}
Using Eqs. (\ref{cc_eqr1}-\ref{cc_eqr2}) in Eq.(\ref{cc_eq}), we get 
the CC equation in the form
\begin{eqnarray}
   && \langle\Phi^{pq}_{vw}|\bar H_{\rm N} + 
      \{\contraction{\bar}{H}{_{\rm N}}{S}\bar H_{\rm N}S\} + 
      \frac{1}{2}\{\contraction{\bar}{H}{_{\rm N}}{S}
      \contraction[1.5ex]{\bar}{H}{_{\rm N}S}{S}\bar H_{\rm N}SS\}- 
      \contraction{}{S}{}{H}SH_{\rm eff} \nonumber \\
   && -\frac{1}{2}\{\contraction{S}{S}{}{H}
      \contraction[1.5ex]{}{S}{S}{H}SSH_{\rm eff}\}|\Phi_{vw}\rangle = 0,
  \label{cc_eq1}
\end{eqnarray}
where $H_{\rm eff} = H_0 + W$, is the effective Hamiltonian.
The form of the effective Hamiltonian 
$H_{\rm eff}$ is close, no free lines or only valence lines as free lines,
therefore Eq.(\ref{cc_eq1}) can be written as
\begin{eqnarray}
    \langle\Phi^{pq}_{vw}| \bar H_{\rm N} +
   \{\contraction{\bar}{H}{_{\rm N}}{S}\bar H_{\rm N}S\} +
   \frac{1}{2}\{\contraction{\bar}{H}{_{\rm N}}{S}\bar H_{\rm N}S^2\}
    |\Phi_{vw}\rangle = \nonumber \\ 
    H_{\rm eff}\langle\Phi^{pq}_{vw}|S + \frac{1}{2}S^2|\Phi_{vw}\rangle.
  \label{cc_eq2}
\end{eqnarray}
This is identical to Eq.(\ref{ccsd_2v}), which is obtained from the
eigen value equation of the Dirac-Coulomb Hamiltonian with the exponential 
ansatz.


\subsection{Diagonalization of $H_{\rm eff}$}

  In the single reference calculations the mapping from 
reference state to exact state is simple, and straight forward. The exact 
state, as given in Eq. (\ref{exact_state}), is the transformation of 
reference state $|\Phi_{vw}\rangle$ with $\Omega$. It is not so 
simple with multi-reference model spaces. The model space then encompasses
a set of determinantal states  $\{|\Phi_{v_\alpha w_\beta}\rangle  \}\in P$, 
however, each state by themselves are not the reference states. The CC 
equation in Eq. (\ref{cc_eq2}) is then modified to
\begin{eqnarray}
   && \langle\Phi^{pq}_{v_\alpha w_\beta}|\bar H_{\rm N} + 
      \{\contraction{\bar}{H}{_{\rm N}}{S}\bar H_{\rm N}S\} + 
      \frac{1}{2}\{\contraction{\bar}{H}{_{\rm N}}{S}
      \contraction[1.5ex]{\bar}{H}{_{\rm N}S}{S}\bar H_{\rm N}SS\}
      |\Phi_{v_\alpha w_\beta}\rangle
           \nonumber \\
   && - \sum_{\gamma,\delta}\langle\Phi^{pq}_{v_\alpha w_\beta}|
      \contraction[1.5ex]{}{S}{+ \frac{1}{2}S^2\phigd\gdphi }{W}
      \contraction{S+\frac{1}{2}}{S}{^2\phigd\gdphi }{W}
      S + \frac{1}{2}S^2\phigd\gdphi W
      |\Phi_{v_\alpha w_\beta}\rangle = 0, \;\;\;\;\;\; 
  \label{cc_eq3}
\end{eqnarray}
where, the sum over $\delta$ and $\gamma$ spans all the determinantal 
states within $P$. This is the working equation of multi-reference 
two-valence CCT with CMS. The  last term require careful consideration while
implementing and as we mentioned earlier, folded diagrams arise from this
term. 

  The wave operator $\Omega$ is defined once the CC equations are solved, but
the model functions are not yet defined. Next step of the calculation is then 
to evaluate the matrix elements of the effective Hamiltonian
\begin{equation}
   H_{\rm eff} (v, w; x, y) = 
    \langle\Phi_{vw}|H_0  
    + V\Omega|\Phi_{xy}\rangle .
\end{equation}
The $H_{\rm eff}$ matrix is non symmetric as $\Omega$ operates on the ket 
state and after diagonalization, one gets a biorthogonal set of eigen states
$|\Psi_i^0\rangle $. These are the model functions of the 
multi-reference CC, the exact state is then
\begin{equation}
   |\Psi_i\rangle = \Omega|\Psi_i^0\rangle,
\end{equation}
and the eigen value equation is 
\begin{equation}
   H_{\rm eff} |\Psi_i^0\rangle = E_i |\Psi_i^0\rangle .
  \label{2v_eigen}
\end{equation}
The eigenstates in general are of the form
\begin{equation}
   |\Psi_i^0\rangle  = \sum_{\alpha \beta} c_{\alpha\beta}^i
                            |\Phi_{v_\alpha w_\beta}\rangle ,
\end{equation}
where $c_{\alpha\beta}^i$ are the coefficients of the linear combination 
or eigen vector elements of $H_{\rm eff}$.


\section{Incomplete model space} 

  Incomplete model space (IMS) consists of a restricted number of 
configurations from the CMS. Remaining configurations are part of the 
orthogonal space. Outcome of such a model space is, the clean separation of 
internal and external cluster amplitudes is no longer true. Further more, the 
subsystem embedding condition is violated. For example, cluster operators which
are external in one-valence Hilbert space may no longer be so in the 
two-valence Hilbert space. The intermediate normalization Eq. (\ref{int_norm}) 
is then, in general, not applicable
\begin{equation}
  |\Psi_i^0\rangle \neq  P\Omega|\Psi_i^0\rangle .
\end{equation}
Following which, the $H_{\rm eff}$ is not guaranteed to be operational only
within the model space, it may as well connect a state in $P$ to 
a state in $Q$. Where as the obvious advantage of defining $H_{\rm eff}$
is to work within the model space and incorporate the effects of orthogonal
space in an effective way. Restoring the operational space of $H_{\rm eff}$ to
$P$ requires a set of constraint equations \cite{Mukherjee-86} and  a previous
work reported the implementation of particle-hole sectors \cite{Hughes-93}
in relativistic CC calculations. However, all the good virtues of CMS, in the 
context of Fock-space CCT, are applicable when the model space is 
quasi-complete. For a lucid description of what constitutes a quasi-complete
model space refer \cite{Lindgren-85ps,Lindgren-87}.

\begin{figure}[h]
\begin{center} 
  \includegraphics[width = 7.2cm]{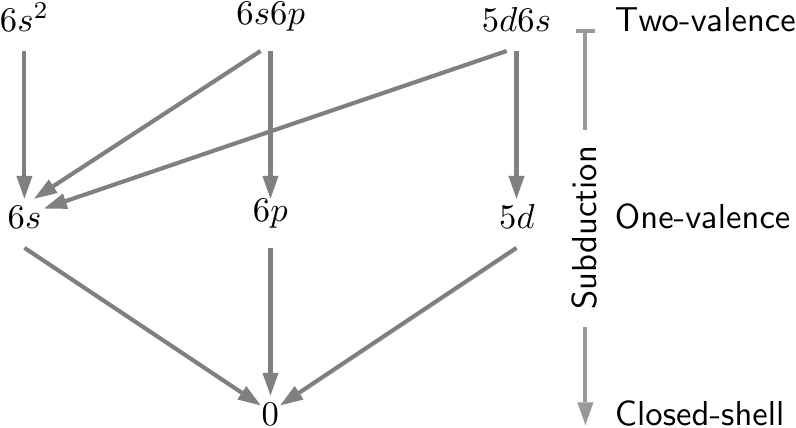} 
  \caption{Incomplete model space of Yb two-valence calculations. Arrows 
           indicate the {\em subduction} to lower valence sectors and 
           respective model spaces.
          }
  \label{yb_incomp}
\end{center}
\end{figure}

  Like in CMS, as a specific example consider the low-lying states of Yb.
An ideal incomplete model space would consist of the configurations $6s^2$, 
$6s6p$ and $5d6s$. Model space would then encompass all the levels important
to ongoing precision experiments: $6s^2~(^1S_0)$, $6s6p~(^3P_J)$, 
$5d6s~(^3D_J)$ and $6s6p~(^1P_1)$. Obvious advantage in such a selection of 
model space is isolation, as evident in Fig. \ref{yb_elevel}, 
from the potential {\em intruder} states $6p7p~(^3P_J)$, $6s7s~(^3S_1)$ and 
$6s7s~(^1S_0)$. Here, we can apply subduction process to check if the model
space considered is quasi-complete and is shown in  Fig. \ref{yb_incomp}. 
Initial stage is the two valence model space consisting
of $6s^2$, $6s6p$ and $5d6s$. Removal of one electron from each of the 
configurations leads to a configuration in one-valence model space ( $6s$, 
$6p$ and $5d$). Finally, removal of another electron gives the closed-shell
model space. All the configurations obtained in the subduction are part
of respective model spaces. This is a requirement of quasi-complete model
space and necessary condition for separation of internal and external 
excitations.


\section{Properties calculations}


\subsection{Hyperfine structure constants}

  The HFS constants of an atom are the parameters which measure further
splitting of fine structure levels. It arises from the interaction of 
electromagnetic moments of the nucleus with the electromagnetic field of the 
atomic electrons \cite{Charles-55}. The general form of the hyperfine 
interaction Hamiltonian is 
\begin{equation}
  H_{\rm hfs} = \sum_i\sum_{k, q}(-1)^q t^k_q(\hat {\mathbf r}_i) T^k_{-q},
  \label{hfs_ham}
\end{equation}
where $t^k_q(\mathbf{r})$ and $T^k_{q}$ are irreducible tensor operators of rank
$k$ effective in the electron and nuclear spaces respectively.
For the magnetic dipole interaction ($k = 1$), the explicit form of the 
tensor operators are
\begin{eqnarray}
  t^1_q(\mathbf{r}) & = &\frac{-i\sqrt{2}[{\bm \alpha}\cdot\mathbf{C}_1
                        (\hat {\mathbf r})]_q} {cr^2},
                         \nonumber \\ 
  T^1_q &= &\mu_q.    
  \label{hfs_magnetic}
\end{eqnarray}
Here, $\mathbf{C}_1(\hat {\mathbf r})$ is a rank one tensor operator in electron
space and $\mu _q$ is a component of $\bm \mu$, the nuclear magnetic moment
operator. The HFS constants are the expectation value of $H_{\rm hfs}$ and 
the magnetic dipole HFS constant is then
\begin{equation}
  a = \frac{\langle \Psi_i|H_{\rm hfs}|\Psi_i\rangle}
           {\langle \Psi_i|\Psi_i\rangle}.
  \label{hfs_mdipole}
\end{equation}
Where, $|\Psi^i\rangle$ is the exact wave function expressed in 
Eq.(\ref{onev_exact}), using coupled-cluster theory. The denominator 
$ \langle \Psi_i|\Psi_i\rangle$ is the normalization factor and it is not be
confused with intermediate normalization Eq. (\ref{int_norm}). The later 
defines the relation between the reference state and the exact state. And it 
does not determine the normalization of the exact state.


\subsection{HFS constant in one-valence sector}

  Once the CC equations and cluster amplitudes are known, the atomic 
properties are calculated with the exact atomic states so obtained. It is
expectation for dynamical variables and matrix element for transition 
amplitudes. From the CC wave function of one valence systems in 
Eq.(\ref{onev_exact}), the expectation of $H_{\rm hfs}$ is 
\begin{equation}
  \langle \Psi_v|H_{\rm hfs}|\Psi_v \rangle = 
       \langle \Phi_v| \tilde H_{\rm hfs} + 2 S^\dagger \tilde H_{\rm hfs} + 
       S^\dagger \tilde H_{\rm hfs} S |\Phi_v\rangle ,
  \label{hfs_num}
\end{equation}
where, $\tilde H_{\rm hfs} = e{^T}^\dagger H_{\rm hfs} e^T$ is the dressed
operator. The factor of two in the second term on the right hand side accounts
for $\tilde H_{\rm hfs} S $ as
$S^\dagger \tilde H_{\rm hfs} = \tilde H_{\rm hfs} S $. An expansion of
$\tilde H_{\rm hfs}$ ideal for an order wise calculations is
\begin{equation}
  \tilde H_{\rm hfs} = H_{\rm hfs} e^T + \sum_{n = 1}^\infty \frac{1}{n!}
                 \left ( T^\dagger \right )^n H_{\rm hfs} e^T.
  \label{A_tilde}
\end{equation}
The normalization factor, denominator in Eq.(\ref{hfs_mdipole}), in terms of
coupled-cluster wave function is
\begin{equation}
  \langle \Psi_v |\Psi_v \rangle = 
      \langle \Phi_v|\left (1 + S^\dagger\right ) e{^T}^\dagger e^T
                     \left ( 1 +  S\right )|\Phi_v\rangle.
\end{equation}
Note that dressed operator $ \tilde H_{\rm hfs}$ and $ e{^T}^\dagger e^T$
in the normalization factor are non terminating series. In a recent work, we 
demonstrated a scheme to include a class of diagrams to all order in $T$ 
iteratively for properties calculations. With the method we calculated the 
magnetic dipole HFS constant of the singly ionized 
alkaline-Earth metals \cite{Mani-10}. Based on the extensive calculations 
reported in ref. \cite{Mani-10}, we conclude terms higher than quadratic in 
$T$ contribute less than $0.1\%$ to the HFS constants. So 
there are no compromises on important physics when $\widetilde H_{\rm hfs}$, 
Eq.(\ref{hfs_truncate}), is truncated after the second-order in $T$. However, 
there are enormous gains in computing resources and simplification of the 
procedure with the iterative scheme. Here, we shall not dwell on the iterative 
scheme, interested readers may refer to ref. \cite{Mani-10} for more details.


\subsection{HFS constant in two-valence systems}

From the CC wave functions of two-valence systems defined in 
Eq.({\ref{twov_exact}}), we get
\begin{eqnarray}
  \langle\Psi_i|H_{\rm hfs}|\Psi_i\rangle&=& \sum_{j,k}{c^i_j}^*c^i_k
     \left[\langle\Phi_j|\tilde H_{\rm hfs}+ \tilde H_{\rm hfs}
     \left(S + \frac{1}{2}S^2 \right ) 
               \right. \nonumber \\
  &&\left.\ + \left(S + \frac{1}{2}S^2 \right )^\dagger\tilde H_{\rm hfs}
    + \left(S + \frac{1}{2}S^2 \right )^\dagger
               \right. \nonumber \\
  &&\left.\; \tilde H_{\rm hfs}\left(S + \frac{1}{2}S^2 \right )
    |\Phi_k\rangle \right].
  \label{hfs_cc}
\end{eqnarray}
Where to shorten the notations we have replaced the valence indexes in the 
two-valence states $v_{\alpha}w_{\beta}$ ($v_{\delta}w_{\gamma}$ ) with $j$ 
($k$). This is the CC expression to calculate the HFS constants of two-valence 
electron atoms. The operator $\tilde H_{\rm hfs}$ is, as defined earlier,
the dressed HFS interaction Hamiltonian. As discussed in the one-valence case, 
comprehensive inclusion of all order of $T$ is beyond the scope of current 
theories  and computational resources. Hence, for the two-valence sector
we consider up to quadratic terms of $T$ in $\tilde H_{\rm hfs}$, approximately 
\begin{equation}
  \widetilde H_{\rm hfs} \approx H_{\rm hfs}+H_{\rm hfs}T 
          + T^\dagger H_{\rm hfs} + T^\dagger H_{\rm hfs}T.
  \label{hfs_truncate}
\end{equation}

\begin{figure}[h]
\begin{center}
  \includegraphics[width = 8.2cm]{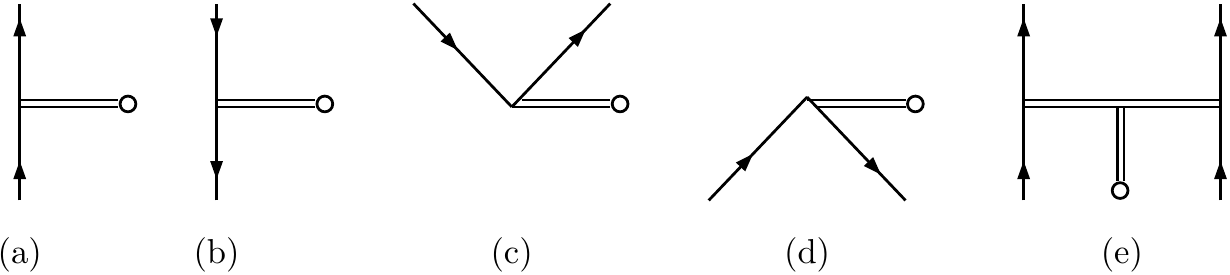}
  \caption{Representation of effective one- and two-body dressed properties
           operators.}
  \label{hfs_otbdy}
\end{center}
\end{figure}

 To compute $\tilde H_{\rm hfs} $ of the two-valence sector,
we borrow the concept of effective one- and two-body operators from our 
previous work ref \cite{Mani-10}. Diagrammatic representation of the effective
operators are as shown in Fig. \ref{hfs_otbdy}.
It is important to note that the two-body effective operator, shown in 
Fig. \ref{hfs_otbdy}(e), arises from the last term in Eq. (\ref{hfs_truncate}). 
And, since it has two-orders of $T$, the actual hyperfine diagrams obtained 
after contraction with $S$ may have negligible contributions. 
For this reason we shall not elaborate on the HFS diagrams arising from the 
dressed two-body effective properties operator. However, we do incorporate
these diagrams in the calculations and mention the contributions 
in the results.
The diagrams of the HFS constant in the two-valence sector are grouped into
different categories. First based on the cluster operators in the expression
and later, in terms of the number of core, valence and virtual orbitals. 
Specific terms and groups of diagrams are discussed in this work.


\subsubsection{Effective one-body operator}
\label{hfs-one-body}

 There are four diagrams from $\widetilde H_{\rm hfs}$ which
has non-zero contribution. These are the two-valence diagrams from
$\contraction{}{T}{_2^{\dagger}}{T}T_2^{\dagger}T_2   $ with the 
bare hyperfine interaction $h_{\rm hfs}$ inserted to all the possible orbital
lines. Contributions from these diagrams is expected to be very small as 
$S$  are not a part of the diagrams. 
\begin{figure}[h]
\begin{center}
  \includegraphics[width = 8.0cm]{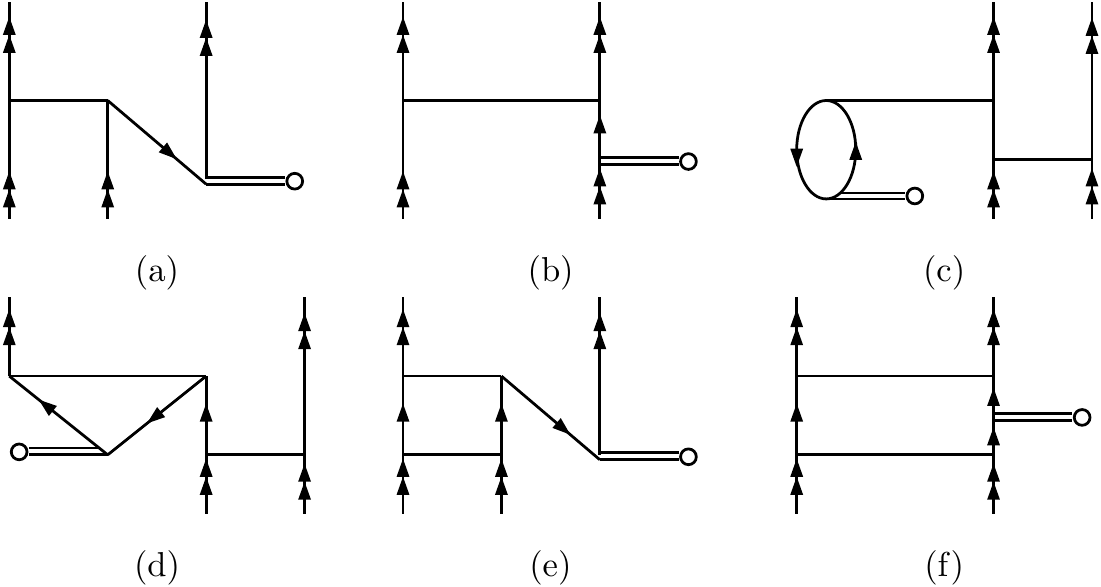}
  \caption{Hyperfine diagrams contributing to the terms,
           ${S^{(1)}_2}^\dagger\widetilde H_{\rm hfs}$ (diagram $(a)$),
           ${S^{(2)}}^\dagger\widetilde H_{\rm hfs}$ (diagram $(b)$),
           ${S^{(1)}_2}^\dagger\widetilde H_{\rm hfs}S^{(2)}$
           (diagrams form $(c)$ to $(e)$),
           and ${S^{(2)}}^\dagger\widetilde H_{\rm hfs}S^{(2)}$
           (diagram $(f)$).}
  \label{hfs_sdas}
\end{center}
\end{figure}

  Diagrammatically, $\widetilde H_{\rm hfs}S^{(1)}$ and 
${S^{(1)}}^\dagger\widetilde H_{\rm hfs}$
each have one diagram and these arise when $S^{(1)}$ is contracted with
the one-body effective operators: $S^{(1)}_2$ with diagram in 
Fig. \ref{hfs_otbdy}(d), and ${S^{(1)}_2}^\dagger$ with diagram in
Fig. \ref{hfs_otbdy}(c). The diagram arising from 
${S^{(1)}_2}^\dagger\widetilde H_{\rm hfs}$ is as shown  in 
Fig. \ref{hfs_sdas}(a). Time reversed version of the same diagram correspond
to $\widetilde H_{\rm hfs}S^{(1)}_2$, however, this is not shown in figure.
The contributions from  $\widetilde H_{\rm hfs}S^{(1)}$ and 
${S^{(1)}}^\dagger\widetilde H_{\rm hfs}$ are large as these are only first 
order in $S$. Further more, $H_{\rm hfs} $ is one-body interaction and 
hence, one-body effective interaction are dominant. 

  The terms $\widetilde H_{\rm hfs}S^{(2)}$ and 
${S^{(2)}}^\dagger\widetilde H_{\rm hfs}$ each have one diagram and These 
arise from the contraction of $S^{(2)}_2 $ with the one-body effective 
operator  of $\widetilde H_{\rm hfs} $ shown in diagram 
Fig.~\ref{hfs_otbdy}(a). The diagram from 
${S^{(2)}}^\dagger\widetilde H_{\rm hfs}$ is shown in 
Fig. \ref{hfs_sdas}(b). Like in the previous case, the time reversed diagram
arise from $\widetilde H_{\rm hfs}S^{(2)}$ and is not shown in the figure. One 
can expect these terms to constitute the leading order as these
are the lowest order terms with $S_2^{(2)}$. Rationale for such an anticipation
is, in general, the magnitudes of $S_2^{(2)}$ are larger than $S_2^{(1)}$ 
and $T$ operators.

\begin{figure}[h]
\begin{center}
  \includegraphics[width = 8.2cm]{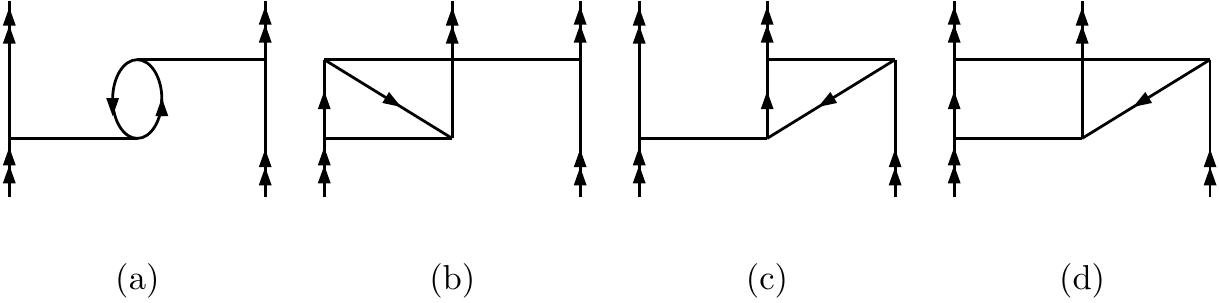}
  \caption{Diagrams arising in contraction of $S{^{(1)}_2}^\dagger$
           with $S^{(1)}_2$.}
  \label{s1ds1}
\end{center}
\end{figure}


\subsubsection{$S^\dagger\widetilde H_{\rm hfs}S$}

\begin{figure}[h]
\begin{center}
  \includegraphics[width = 8.2cm]{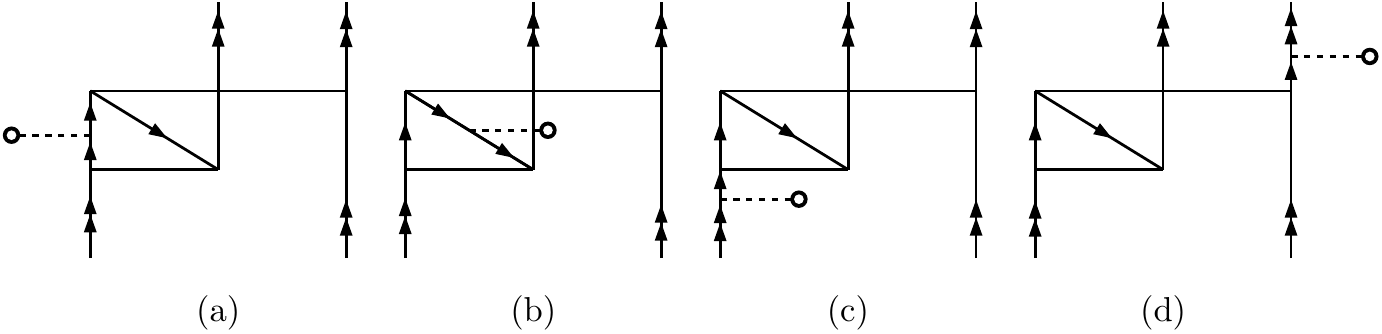}
  \caption{Hyperfine diagrams obtained after inserting the Hyperfine
           interaction operator in diagram $(b)$ of Fig.(\ref{s1ds1}).}
  \label{s1ds1_hfs}
\end{center}
\end{figure}

The leading term in Eq. (\ref{hfs_cc}) which is quadratic in $S$ is 
\begin{eqnarray}
  \!\!\!\!\!\!S^\dagger\tilde H_{\rm hfs}S &= &
  {S^{(1)}}^\dagger\tilde H_{\rm hfs}S^{(1)} + 
  \left [ {S^{(2)}}^\dagger\tilde H_{\rm hfs}S^{(1)} + {\rm c.c.} \right ] 
            \nonumber  \\
  && + {S^{(2)}}^\dagger\tilde H_{\rm hfs}S^{(2)}.
\end{eqnarray}
where, c.c. represents complex conjugation. We now discuss the diagrams arising
from each of these terms. There are sixteen diagram arising from  
${S^{(1)}}^\dagger\widetilde H_{\rm hfs}S^{(1)}$ and topologically, these 
are the effective one-body diagrams Fig.~\ref{hfs_otbdy}(a-b) sandwiched 
between ${S^{(1)}}^\dagger$ and $S^{(1)}$. To examine the diagrams in more 
detail, all the diagrams (four in all) from the contraction
$\contraction{}{S}{{^{(1)}_2}^\dagger}{S}{S^{(1)}_2}^\dagger S^{(1)}_2$ are
shown in in Fig.~\ref{s1ds1}. To each of the diagrams in Fig.~\ref{s1ds1} 
the effective one-body operator can be inserted in four ways. As an example,
consider the diagram in Fig. \ref{s1ds1}(b), all the four diagrams after 
inserting the effective one-body operator are shown in Fig.~\ref{s1ds1_hfs}.

  There are five diagrams from 
${S^{(1)}}^\dagger\widetilde H_{\rm hfs}S^{(2)}$. 
These arise from the  contraction of ${S^{(1)}}^\dagger$ with 
$S^{(2)}$ through one-body operators in Fig. \ref{hfs_otbdy}(a) and (c).
The diagrams from ${S^{(1)}_2}^\dagger\widetilde H_{\rm hfs}S^{(2)}$ 
are as shown in Fig.~\ref{hfs_sdas}(c-e). However, the diagrams from
${S^{(1)}_1}^\dagger$ are not shown. Identical number of diagrams arise from
${S^{(2)}}^\dagger\widetilde H_{\rm hfs}S^{(1)}$. The effective diagram
in this case are Fig. \ref{hfs_otbdy}(a) and (d).

 Finally, only one diagram arises from the last term, 
${S^{(2)}}^\dagger\widetilde H_{\rm hfs}S^{(2)}$. This diagram is shown 
Fig. \ref{hfs_sdas}(f). Only Fig.~\ref{hfs_otbdy}(a) is the allowed 
effective one-body operator which contribute to this
term.

\subsection{Electric dipole transition amplitudes}

 Electric dipole is the most dominant electromagnetic multipole in the 
radiative transition of atoms. In majority of the cases, depending on the 
decay channels, it defines the life time of an excited state. Theoretically, 
the relevant quantity is the reduced matrix element of the dipole operator 
$\mathbf{D}$ between the initial and final states $|\Psi_i\rangle$ and 
$|\Psi_f\rangle$, respectively. The two states are opposite in parity as 
$\mathbf{D}$ is an odd parity operator. The expression of the reduced matrix 
element is  
\begin{equation}
  D_{if} = \frac{\langle \Psi_f||D||\Psi_i\rangle}
           {\sqrt{\langle \Psi_f|\Psi_f\rangle\langle\Psi_i |\Psi_i\rangle}}.
  \label{d_if}
\end{equation}
Once the reduced matrix elements are evaluated, the actual matrix elements 
of the specific states are calculated from the Wigner-Eckert theorem.
Here, we need to make a finer distinction of the wave operator
in the valence universal or Fock-space CCT. As $H^{\rm DC}$ commutes with 
parity, so does the wave operator $\Omega$ and we can consider $\Omega$ as
\begin{equation}
  \Omega = \Omega ^+ + \Omega ^-.
\end{equation} 
Where $\Omega ^+ $ and $ \Omega ^-$ operates on the even and odd parity 
reference states within the model space. The separation of these two components
follows naturally from the parity selection rules imposed on the cluster
amplitudes. However, this ought to be handled with care as complications arise 
in the calculations of peturbed cluster amplitudes. Where there is a second 
perturbation, besides the residual Coulomb interaction, which is parity
odd. We may rewrite Eq. (\ref{d_if}) more precisely as
\begin{equation}
  D_{if} = \frac{\langle \Psi_f^0||{\Omega^{\mp}}^\dagger D\Omega^{\pm}
           ||\Psi_i^0\rangle}
           {\sqrt{\langle \Psi_f|\Psi_f\rangle\langle\Psi_i |\Psi_i\rangle}}.
\end{equation}
In the one-valence sector, the reduced matrix element of $D$ is 
\begin{equation}
  \langle \Psi_w||D||\Psi_v \rangle = 
       \langle \Phi_v|| \tilde D + S^\dagger \tilde D + \tilde D S +
       S^\dagger \tilde D S ||\Phi_v\rangle ,
  \label{dip_1v}
\end{equation}
where, $|\Psi_v \rangle$ and $|\Psi_w \rangle$ are the initial and final 
states in terms of the valence states. Though the expressions are similar to 
Eq.(\ref{hfs_num}), there are two important differences. Unlike the 
HFS energy splitting expression $S^\dagger \tilde D \neq \tilde DS$, this is 
because the initial and final states are different. Same set of 
properties diagrams in HFS calculations, after modifications to account for
the two key differences,  are then adopted to compute reduced $\bm{D}$ matrix
elements with CC wave functions.

 After a similar modification, like in Eq. (\ref{hfs_cc}), for the two-valence 
systems 
\begin{eqnarray}
  \langle\Psi_f||D||\Psi_i\rangle&=& \sum_{j,k}{c^f_j}^*c^i_k
     \left[\langle\Phi_j|\tilde D + \tilde D \left(S + \frac{1}{2}S^2 \right ) 
               \right. \nonumber \\
  &&\left.\ + \left(S + \frac{1}{2}S^2 \right )^\dagger\tilde D
    + \left(S + \frac{1}{2}S^2 \right )^\dagger
               \right. \nonumber \\
  &&\left.\; \tilde D\left(S + \frac{1}{2}S^2 \right )
    |\Phi_k\rangle \right].
  \label{dip_2v}
\end{eqnarray}
Where the notations and terms are the same as in the two-valence HFS case. 
However, the two key differences mentioned earlier still hold true. Like
in HFS
\begin{equation}
  \widetilde D \approx D + DT + T^\dagger D + T^\dagger DT.
  \label{dip_truncate}
\end{equation}
We then proceed like in HFS and calculate the effective diagrams, both one- and 
two-body. These are then contracted with the cluster operators and we
evaluate the transition matrix.


\section{Numerical details}

  Coupled-cluster theory, though powerful is computationally intensive and 
implementation is non trivial. The large number of unknowns and equations
demand special attention to all aspects of computations. Right from  the
initial stage of identifying and calculating the cluster diagrams, to the 
final stages of solving the CC equations and computing properties from the 
CC wave functions. Here, we give concise description of what we consider 
absolutely essential, theoretical and computational aspects for atomic 
coupled-cluster calculations. This choice is entirely based on our experience 
of developing and implementing CCT spanning closed-shell, one- and two-valence 
systems. Besides the CC wave-function calculations, we have also proposed, 
developed and implemented methods to compute properties from CC wave functions.
The selected issues addressed are provided with the anticipation that
interested researchers shall find these details valuable. And facilitate
adopting CCT for atomic many-body computations with minimal effort.


\subsection{Orbitals and basis functions}

 Results presented in this paper are based on the Dirac-Coulomb 
Hamiltonian $H^{\rm DC}$ given in Eq. (\ref{dchamil}). It incorporates 
relativity at the single particle level accurately. And, as the name 
indicates, the Coulomb interactions between the electrons. For the nuclear 
potential $V_N(r)$, we consider the  finite size Fermi density distribution
\begin{equation}
  \rho_{\rm nuc}(r) = \frac{\rho_0}{1 + e^{(r-c)/a} },
\end{equation}
here, $a = t 4\ln 3$. The parameter $c$ is the half-charge radius, that is
$\rho_{\rm nuc}(c)=\rho_0/2$ and $t$ is the skin thickness. At the single 
particle level, the spin orbitals are of the form
\begin{equation}
  \psi_{n\kappa m}(\mathbf{r})=\frac{1}{r}
  \left(\begin{array}{r}
            P_{n\kappa}(r)\chi_{\kappa m}(\mathbf{r}/r)\\
           iQ_{n\kappa}(r)\chi_{-\kappa m}(\mathbf{r}/r)
       \end{array}\right),
  \label{spin-orbital}
\end{equation}
where $P_{n\kappa}(r)$ and $Q_{n\kappa}(r)$ are the large and small component
radial wave functions, $\kappa$ is the relativistic total angular momentum
quantum number and $\chi_{\kappa m}(\mathbf{r}/r)$ are the spin or spherical
harmonics. One representation of the radial components is to define these
as linear combination of Gaussian like functions and are referred as
Gaussian type orbitals (GTOs). Then, the large and small
components \cite{Mohanty-89,Chaudhuri-99} are
\begin{eqnarray}
   P_{n\kappa}(r) = \sum_p C^L_{\kappa p} g^L_{\kappa p}(r),  \nonumber \\
   Q_{n\kappa}(r) = \sum_p C^S_{\kappa p} g^S_{\kappa p}(r).
\end{eqnarray}
The index $p$ varies over the number of the basis functions.
For large component we choose
\begin{equation}
  g^L_{\kappa p}(r) = C^L_{\kappa i} r^{n_\kappa} e^{-\alpha_p r^2},
\end{equation}
here $n_\kappa$ is an integer. Similarly, the small component is derived from 
the large components using kinetic balance condition \cite{Stanton-84}. The
exponents in the above expression follow the general relation
\begin{equation}
  \alpha_p = \alpha_0 \beta^{p-1}.
  \label{param_gto}
\end{equation}
The parameters $\alpha_0$ and $\beta$ are optimized for each of the ions or 
neutral atoms to provide a good description of the properties. In our case the 
optimization criteria are to reproduce the numerical result of the self
consisten field (SCF) energy and orbital energies. 

  From Eq.(\ref{spin-orbital}) the reduced matrix element of the magnetic 
hyperfine operator between two spin orbitals , $v'$ and $v$, is
\begin{eqnarray}
  \langle v'||t^1||v\rangle &=& -(\kappa_v + \kappa_{v'})
  \langle -\kappa_{v'}||C^1||\kappa_v \rangle \times \nonumber \\
  &&\int^\infty_0 \frac{dr}{r^2}(P_{n_{v'}\kappa_{v'}} Q_{n_v\kappa_v}
                                  + Q_{n_{v'}\kappa_{v'}} P_{n_v\kappa_v}).
  \label{hfs_matrix}
\end{eqnarray}
A detailed derivation is given in Ref. \cite{Johnson-07}.

  For the alkaline Earth metal atoms Sr and Ba as well as Yb, we 
use $V^{N-2}$ orbitals for the calculations. This is equivalent
to calculating the spin orbitals from the single particle eigenvalue equations
of the doubly ionized atoms, namely Sr$^{2+}$, Ba$^{2+}$ and Yb$^{2+}$. The 
single particle basis sets then have few bound states and rest are continuum. 
The basis set is optimizeed such that: single particle energies of the core and 
valence orbitals are in good agreement with the numerical results. For this 
we use GRASP92 \cite{Parpia-96} to generate the numerical results.


\subsection{Orbital subsets}

\begin{figure}[h]
\begin{center}
  \includegraphics[width = 5cm]{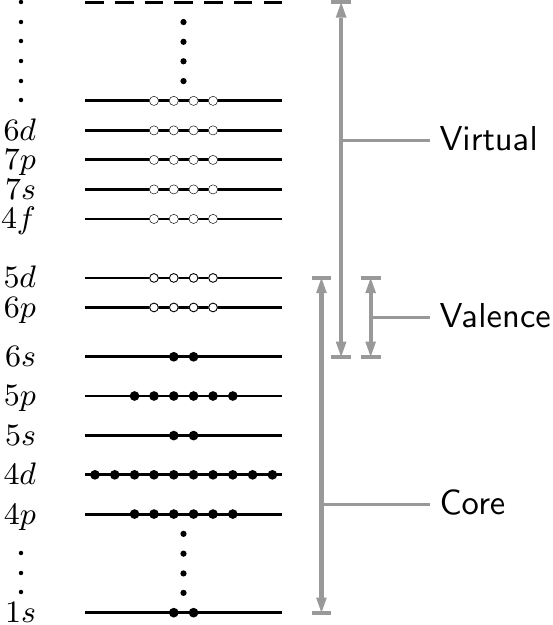}
  \caption{Classification of orbitals into core, valence and virtual subsets. 
           Few orbitals are members of more than one subset.
           }
  \label{yb_basis}
\end{center}
\end{figure}

 Orbitals, the single electron wave functions, in closed-shell systems are
separated into two distinct subsets, core (occupied) and virtual (unoccupied).
Former are shells which are completely filled in the ground state 
determinantal state and later are empty. Distinction is not so straight forward
in open-shell systems. The classification of the orbitals for Yb atom in our 
current calculations is shown in Fig. \ref{yb_basis}. The valence orbitals are 
partially filled in the model functions and are like core orbitals, electrons 
can be excited from the valence shells. 
This is particularly true when considering the valence as particles. 
Consequently, as 
discussed in the next subsection, the closed-shell diagrams  can be modified
to the open shell ones. On the other hand, valence shells can also accommodate
excitations from the core shells. A property typical of shells in the virtual 
space. Hence, in the cluster amplitudes the excited states incorporate the 
valence orbitals as well. The dual character of the valence orbitals can be
adapted for faster diagram evaluations with appropriate rearrangement
of the summation sequence. For example, there is enormous computational
advantage in considering the free orbital lines, in the cluster diagrams,
as the outer most summation. The example given and many other features we 
have developed are more computational in nature and less of physics. We
shall elaborate on these matter in future publications devoted to the 
computational aspects of our work.


\subsection{CC equations}

  In Fock-space CCT, as mentioned earlier, the cluster operators
are generated sector wise in sequence. First, the closed-shell cluster 
amplitudes are computed from the Eq. (\ref{t1_eqn}-\ref{t2_eqn}). Where the
dressed operator $\bar H_{\rm N}$ in the closed-shell CC equations, like in 
Eq. (\ref{cc_eqr1}), is 
\begin{eqnarray}
  \bar H_{\rm N} & = & H_{\rm N} +\{\contraction{}{H_{\rm N}}{}{T}H_{\rm N}T\} +
  \frac{1}{2!}\{\contraction{}{H_{\rm N}}{}{T}
  \contraction[1.5ex]{}{H_{\rm N}}{T}{T}H_{\rm N}TT\} +
  \frac{1}{3!}\{\contraction{}{H_{\rm N}}{}{T}
  \contraction[1.5ex]{}{H_{\rm N}}{T}{T}
  \contraction[2ex]{}{H_{\rm N}}{TT}{T}H_{\rm N}TTT\}    \nonumber \\
  && +\frac{1}{4!}\{\contraction{}{H_{\rm N}}{}{T}
  \contraction[1.5ex]{}{H_{\rm N}}{T}{T}
  \contraction[2ex]{}{H_{\rm N}}{TT}{T}
  \contraction[2.5ex]{}{H_{\rm N}}{TTT}{T}H_{\rm N}TTTT\}.
  \label{hbar_exp}
\end{eqnarray}
The closed shell CC equations as evident from the expression of 
$\bar H_{\rm N}$ are nonlinear equation. In CCSD approximation, we have second 
and fourth order nonlinearities in $T_2$ and $T_1$, respectively. However, 
the working equations can be written in linear form as
\begin{eqnarray}
   A_{11}(T)T_1 +  A_{12}(T)T_2 &=& B_1, \\
   A_{21}(T)T_1 +  A_{22}(T)T_2 &=& B_2.
\end{eqnarray}
Since the original equations are nonlinear equations, the coefficients 
$A_{ij}(T)$ are functions of cluster amplitudes $T$. On the right side 
$B$ are the matrix elements of $H_{\rm N}$. The equations are then solved
iteratively till convergence.
\begin{figure}[h]
\begin{center}
  \includegraphics[width = 8.2cm]{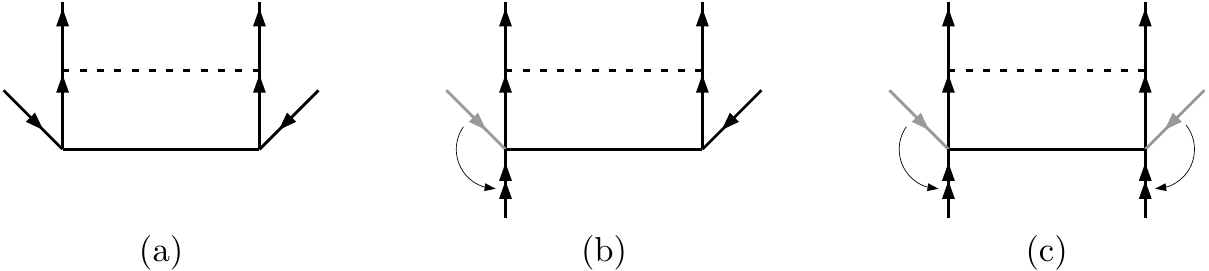}
  \caption{ Conversion from closed-shell cluster operator $T$ diagrams to
            open shell operators $S$.
           }
  \label{ttos1s2}
\end{center}
\end{figure}

  To set up the equations, we evaluate the terms based on diagrammatic 
analysis. There are several diagrams and for example, those arising from linear 
terms are given in ref. \cite{Lindgren-85}. The total number 
of equations scale as $N_v^2N_o^2$, where $N_v$ and $N_o$ are number of 
virtual and occupied orbitals, respectively. For the calculations discussed in 
this paper $N_v\approx 130$ or more and $N_o\approx 20$. The coefficient 
matrix $A$ is non-symmetric and dense, so the number of matrix elements scales
as $\sim N_v^4N_o^4$, which is $\sim 4.6\times 10^{13}$ for typical examples
in the present computations. It is an extremely large matrix and impractical 
to store. In addition, the elements are functions of $T$ and not static. For 
these reasons, the elements of $A$ are calculated on the fly, as and when 
needed. Another complication is, the operations required to generate each 
element of $A$ in the equations scale as $N_v^4N_o^2$. All together, combining 
the number of matrix elements and number of operations, number of binary 
arithmetic operation in each iteration is $O(N_v^8N_o^6)$. Which is indeed 
a very large number for high $Z$ atoms.

 Diagrammatically, to generate the closed-shell equations, we identify all the 
diagrams in the closed-shell CC equations and evaluate the angular 
integrations based on angular momentum diagrams \cite{Lindgren-85}. An example 
diagram is shown in Fig. \ref{ttos1s2}(a), it is the double contraction of 
$V$ with $T_2$ and contributes to the $T_2$ equation. To set up the one-valence 
and two-valence cluster equations we avoid diagrammatic evaluation. Instead, 
the closed-shell diagrams are topologically transformed into one-valence 
diagrams. As shown in Fig. \ref{ttos1s2}(b), one of the core orbital line is 
rotated and transformed into a valence line. Diagrams so obtained are very 
different from Fig. \ref{ttos1s2}(a) in terms of possible contractions. 
However, the results from the angular integration remain unchanged. A similar
procedure is adopted for the two-valence equations as well. With this, a 
careful analysis and evaluation of closed-shell CC equations is the only
requirement to set up the one- and two-valence CC equations. The coupled 
nonlinear and linear equations are solved iteratively. We employ direct 
inversion in the iterated subspace (DIIS) \cite{Pulay-80} for convergence 
acceleration.


\section{Results}


\subsection{Excitation Energies}

  Excitation energies of the one-valence ions Sr$^+$ and Ba$^+$ were reported
in our previous paper \cite{Mani-10}. Here, we report the results of the
calculations in the two-valence sector, the excitation energy of an 
state $nln'l'\;^{(2S+1)}L_J$ , from Eq. (\ref{2v_eigen}), is
\begin{equation}
  \Delta E_{nln'l'\;^{(2S+1)}L_J}= E_{nln'l'\;^{(2S+1)}L_J}-E_{ns^2\;^1S_0}. 
\end{equation}
Where, $E_{ns^2\;^1S_0} $ and $E_{nln'l'\;^{(2S+1)}L_J} $ are the 
ground and excited state eigenvalues of $H_{\rm eff}$ in Eq. (\ref{2v_eigen}).
Vaeck, Godfroid and Hansen \cite{Vaeck-88} had calculated the excitation 
energies and investigated in detail the configuration mixing of atomic Sr with 
multiconfiguration Hartree-Fock theory with special attention on the singlet 
states $^1L$. In particular the states with configurations of the form
$5snp\; ^1P^{\circ}$, $5snd\; ^1D$ and $5snf\; ^1F^{\circ}$, including the 
Rydeberg states. Improved experimental data and prospects of  cooling and 
trapping had spurred further theoretical studies on properties of Sr. 
Important recent theoretical work are by Porsev and collaborators 
\cite{Porsev-01}, and Savukov and Johnson \cite{Savukov-02}. Previous works of 
Eliav, Kaldor and Ishikawa \cite{Eliav-96,Eliav-95} reported the excited 
energy calculations of atomic Ba and Yb using the Fock-space based 
coupled-cluster theory. The other widely used atomic many-body method 
employed is CI-MBPT \cite{Dzuba-96}, based on Dzuba and Ginges \cite{Dzuba-06} 
calculated the excitation energies of Ba. Same method was used by Porsev and 
collaborators to calculate the excitations energies and HFS constants of 
Yb \cite{Porsev-99}.

\begin{table*}[t]
\caption{Two-electron removal energy and the excitation energies calculated
         using relativistic coupled-cluster theory. All values are in 
         atomic units.}
\label{tab-ip-ee}
\begin{ruledtabular}
\begin{tabular}{ccccc}
 State& \multicolumn{2}{c}{Our result}& Other work& Exp result
                                                    {Ref\cite{nist}.}      \\
  \hline                                                                   \\
                       & $E_{vw}$ & EE      & EE   & EE                    \\
  \hline                                                                   \\
      &    & Atomic $^{87}$Sr$;\;$[Kr]$5s^2$&     &                        \\
$5s^2\;^1S_0$ & $-0.61939$ & $0.0    $ & $0.0$ & $0.0$                     \\
$5s5p\;^3P_0$ & $-0.55169$ & $0.06771$ & $0.06566^{\rm a}$
                                                    &$0.06524$             \\
$5s5p\;^3P_1$ & $-0.55170$ & $0.06768$ & $0.06651^{\rm a}$,
                                         $0.06871^{\rm b}$ & $0.06609$     \\
$5s5p\;^3P_2$ & $-0.55203$ & $0.06736$ & $0.06833^{\rm a}$ & $0.06788$     \\
$5s4d\;^3D_1$ & $-0.53551$ & $0.08388$ & $0.08230^{\rm a}$ & $0.08274$     \\
$5s4d\;^3D_2$ & $-0.53478$ & $0.08461$ & $0.08260^{\rm a}$ & $0.08301$     \\
$5s4d\;^3D_3$ & $-0.53397$ & $0.08542$ & $0.08312^{\rm a}$ & $0.08347$     \\
$5s4d\;^1D_2$ & $-0.52594$ & $0.09345$ & $0.09210^{\rm a}$,
                                         $0.11477^{\rm c}$ & $0.09181$     \\
$5s4d\;^1P_1$ & $-0.51283$ & $0.10656$ & $0.09851^{\rm a}$,
                                         $0.10015^{\rm b}$,
                                         $0.10730^{\rm c}$ & $0.09887$     \\
                      \\
       &   & Atomic $^{137}$Ba$;\;$[Xe]$6s^2$ &   &                        \\ 
$6s^2\;^1S_0$ & $-0.56439$ & $0.0    $ & $0.0    $  &$0.0$                 \\
$6s5d\;^3D_1$ & $-0.52303$ & $0.04136$ & $0.04211^{\rm d}$,
                                         $0.04106^{\rm e}$,
                                         $0.04119^{\rm f}$ & $0.04116$     \\
$6s5d\;^3D_2$ & $-0.52170$ & $0.04269$ & $0.04296^{\rm d}$,
                                         $0.04193^{\rm e}$,
                                         $0.04200^{\rm f}$ & $0.04199$     \\
$6s5d\;^3D_3$ & $-0.51960$ & $0.04479$ & $0.04473^{\rm d}$,
                                         $0.04375^{\rm e}$,
                                         $0.04366^{\rm f}$ & $0.04375$     \\
$6s5d\;^1D_2$ & $-0.51030$ & $0.05409$ & $0.05395^{\rm d}$,
                                         $0.05197^{\rm e}$,
                                         $0.05298^{\rm f}$ & $0.05192$     \\
$6s6p\;^3P_0$ & $-0.50667$ & $0.05772$ & $0.05697^{\rm d}$,
                                         $0.05575^{\rm e}$,
                                         $0.05591^{\rm f}$ & $0.05589$     \\
$6s6p\;^3P_1$ & $-0.50540$ & $0.05899$ & $0.05869^{\rm d}$,
                                         $0.05742^{\rm e}$,
                                         $0.05758^{\rm f}$ & $0.05758$     \\
$6s6p\;^3P_2$ & $-0.50311$ & $0.06128$ & $0.06284^{\rm d}$,
                                         $0.06147^{\rm e}$,
                                         $0.06159^{\rm f}$ & $0.06158$     \\
$6s6p\;^1P_1$ & $-0.47291$ & $0.09148$ & $0.08409^{\rm d}$,
                                         $0.08256^{\rm e}$,
                                         $0.08125^{\rm f}$ & $0.08229$     \\
                           \\
       &   & Atomic $^{173}$Yb$;\;$[Xe]$4f^{14}6s^2$ &   &                 \\   
$6s^2\;^1S_0$ & $-0.68083$ & $0.0$ & $0.0$&$0.0$                           \\
$6s6p\;^3P_0$ & $-0.59944$ & $0.08140$ & $0.07909^{\rm g}$,
                                         $0.07874^{\rm h}$,
                                         $0.07877^{\rm i}$& $0.07877$      \\
$6s6p\;^3P_1$ & $-0.59645$ & $0.08439$ & $0.08242^{\rm g}$,
                                         $0.08200^{\rm h}$,
                                         $0.08200^{\rm i}$ & $0.08198$     \\
$6s6p\;^3P_2$ & $-0.58914$ & $0.09170$ & $0.09038^{\rm g}$,
                                         $0.08999^{\rm h}$,
                                         $0.09002^{\rm i}$ & $0.08981$     \\
$6s5d\;^3D_1$ & $-0.56110$ & $0.11973$ & $0.11362^{\rm g}$,
                                         $0.11425^{\rm h}$,
                                         $0.11158^{\rm i}$ & $0.11158$     \\
$6s5d\;^3D_2$ & $-0.55975$ & $0.12109$ & $0.11473^{\rm g}$,
                                         $0.11136^{\rm h}$,
                                         $0.11274^{\rm i}$ & $0.11278$     \\
$6s5d\;^3D_3$ & $-0.55602$ & $0.12481$ & $0.11699^{\rm g}$,
                                         $0.11503^{\rm h}$,
                                         $0.11517^{\rm i}$ & $0.11514$     \\
$6s6p\;^1P_1$ & $-0.55301$ & $0.12782$ & $0.12426^{\rm g}$,
                                         $0.11253^{\rm h}$,
                                         $0.11669^{\rm i}$ & $0.11422$     \\
$6s5d\;^1D_2$ & $-0.54667$ & $0.13416$ & $0.13025^{\rm g}$,
                                         $0.12595^{\rm h}$,
                                         $0.12672^{\rm i}$ & $0.12611$     \\
\end{tabular}
\end{ruledtabular}
\begin{tabbing}
$^{\rm a}$ Reference\cite{Porsev-01}. \;\;\;\= 
                          $^{\rm b}$ Reference \cite{Savukov-02}.    \\
$^{\rm c}$ Reference \cite{Vaeck-88}.     \>
                          $^{\rm d}$ Reference\cite{Eliav-96}.       \\
$^{\rm e}$ Reference\cite{Safronova-09a}.  \>
                          $^{\rm f}$ Reference \cite{Dzuba-06}.      \\
$^{\rm g}$ Reference\cite{Eliav-95}.      \>
                          $^{\rm h}$ Reference\cite{Porsev-99}.      \\
$^{\rm i}$ Reference\cite{Dzuba-10}.      \\
\end{tabbing}
\end{table*}
One reason for choosing the three atoms in our calculations is the 
significant difference in the sequences of $ ns(n-1)d\; ^3D_J$, 
$ ns(n-1)d\; ^1D_2$, $ nsnp\; ^3P_J$ and $ nsnp\; ^1P_1$ levels. As evident 
from Table. \ref{tab-ip-ee}, in Sr the $ 5s4d\; ^{2S+1}D_J$ levels 
lies between $ 5s5p\; ^3P_J$ and $ 5s5p\; ^1P_1$. Whereas the 
$ 6s5d\; ^{2S+1}D_J$ levels are below the $ 6s6p\; ^{2S+1}P_J$ levels
in Ba. The difference in the level structure can be attributed to the 
presence of an additional diffuse shell $4d$. The sequence gets more 
complicated in Yb, $ 6s6p\; ^3P_J$ levels are below $ 6s5d\; ^3D_J$, however, 
the $ 6s6p\; ^1P_1$ lies between $ 6s5d\; ^3D_2$ and $ 6s5d\; ^3D_3$. 
It is to be noted that, the difference between Ba and Yb configurations 
is the presence of $4f$ in the Yb core. And, is the cause for the change in
the level sequence.

The excitation energies obtained from our calculations are reasonably close to 
the other theoretical results for Sr and Ba. However, there is a lack of clear
trend in the differences. For the excitation energies of Sr, our results are 
consistently better than the MCHF results \cite{Vaeck-88}. And, our
result of $5s5p\; ^3P_2$ is closest to the experimental value.  One observation
is, although Porsev and collaborators \cite{Porsev-01}, and Savukov and 
Johnson \cite{Savukov-02} used the same method CI-MBPT, the results from the 
former are consistently better than the later. Possible reason could be the
single particle basis set. The former used a combination of $V^N$, 
$V^{N-1}$ and  $V^{N-2}$ orbitals for the core and valence, and virtuals 
are generated through a recurrent procedure. In the later work, the 
orbital set are B-splines. As described earlier, we use numerical Gaussian
type orbitals for our calculations. 

  Comparison of excitation energies of Ba presents an interesting case. Some
of our results are better than the previous CC results of  Eliav and 
collaborators \cite{Eliav-96}. The results provides a numerical validation of 
our approach. On the other hand, the results of Dzuba and collaborators
\cite{Dzuba-96}, and Safronova and collaborators \cite{Safronova-09a} uses 
similar basis sets but different many-body methods. The former used
CI-MBPT, whereas the later used the recently developed CI plus all order
method \cite{Safronova-09a}. The results from the later are consistently 
better than the former.  

  The Yb excitation energy calculations presents a serious challenge. Earlier
CC calculations of Yb excitation energies  \cite{Eliav-95} could not 
reproduced the experimental sequence and we also encounter the same issue. 
In particular, the $6s6p\; ^1P_1 $ is above the $6s5d\; ^3D_J $ levels,
whereas experimentally it lies between $6s5d\; ^3D_2 $  and $6s5d\; ^3D_3 $.
Sequence in our results is similar and individual values are consistently 
higher than the previous CC calculations. The sequence, however, is correctly 
reproduced in another
calculation with the CI-MBPT method \cite{Porsev-99}, where the basis set 
used is combination of Dirac-Fock orbitals for the core and valence, and 
virtuals are generated through recurrent procedure. The comparison of the 
different results indicate a wide variation in the many-body methods and 
single particle states. In fact, none of the works listed and referred have
a common many-body theory and basis sets. Perhaps, this is an indication of 
the issues which require consistent  efforts to resolve the difficulties of 
precision calculations of two-valence systems.


\subsection{Hyperfine structure constants}

  Hyperfine constants are appropriate atomic properties to inspect the 
accuracy of atomic wave functions in the small radial distances--within 
and close to the nucleus. For the calculations we use the nuclear moments
given in the compilation of Stone \cite{Stone-05} and the values are given
Table. \ref{nuc-mom}.

\begin{table}[h]
\caption{The nuclear spin I, the magnetic moment $\mu$ (in nuclear magneton),
         and the electric quadrupole moment $Q$ (in barn), used in the paper.}
\label{nuc-mom}
\begin{ruledtabular}
\begin{tabular}{cccc}
 Ion        & $I$    & $\mu$     & $Q$       \\
\hline                                       \\
 $^{87}$Sr  & $9/2$&$-1.0936(13)$&$+0.305   $ \\    
 $^{137}$Ba & $3/2$&$+0.93737(2)$&$+0.246(2)$ \\    
 $^{173}$Yb & $5/2$&$-0.648(3)  $&$+2.80(4) $ \\    
\end{tabular}
\end{ruledtabular}
\end{table}


\subsubsection{Hyperfine constant $A$ of Yb$^+$}

  The magnetic dipole HFS constant of $^{173}$Yb$^+$ and the electric 
quadrupole HFS constants of $^{87}$Sr$^+$, $^{137}$Ba$^+$ and $^{173}$Yb$^+$
from our calculations are given in Table. \ref{tab-hfs}. For comparison, the
results from other theoretical studies and experimental data are also given.
Contributions from the specific terms in the CC properties expression of 
HFS are listed in Table. \ref{tab-hfs-comp}. Previous theoretical study by
Martensson \cite{Martensson-94} on the magnetic dipole HFS constant of 
$^{173}$Yb$^+$  is based on the CCT and basis set is obtained from discrete 
spectrum method \cite{Salomonson-89}. In this work, the HFS constant $a$ of 
$6s_{1/2}$, $6p_{1/2}$ and $6p_{3/2}$ are calculated.
\begin{table*}
\caption{The magnetic dipole HFS constant for $^{173}$Yb$^+$ and the 
         electric quadrupole HFS constant for $^{87}$Sr$^+$, $^{137}$Ba$^+$,
          and $^{173}$Yb$^+$ ions. All the values are in MHz.}
\label{tab-hfs}
\begin{ruledtabular}
\begin{tabular}{ccccc}
Ion & state & This work & Other works & Experiment                    \\
\hline
             \\
    &       &  Magnetic dipole HFS constant $A$ & &                   \\
$^{173}$Yb$^+$
      &$6s_{1/2}$ & $-3529.660$ & $-3507^{\rm a}$ & $-3497.5(6)^{\rm a}$,
                                                    $-3508(9)^{\rm c}$     \\
      &$6p_{1/2}$ & $-612.362$  & $-638 ^{\rm a}$ & $-518.2(4)^{\rm a}$,
                                                    $-600^{\rm c}$         \\
      &$6p_{3/2}$ & $-88.973$   & $-107 ^{\rm a}$ & $-$                    \\
      &$5d_{3/2}$ & $-104.479$  & $-110.31^{\rm b}$ & $-$                  \\
      &$5d_{5/2}$ & $22.078$    & $3.47^{\rm b}$    &  $-$                 \\
              \\
      &       &Electric quadrupole HFS constant $B$ & &                    \\
$^{87}$Sr$^+$
      &$5p_{3/2}$ & $84.806$    & $82.655^{\rm d}$, $83.662^{\rm e}$ &
                                                    $88.5(5.4)^{\rm g}$    \\
      &$4d_{3/2}$ & $33.961$    & $35.075^{\rm d}$, $36.051^{\rm e}$,
                                  $39.60^{\rm b}$   &  $-$                 \\
      &$4d_{5/2}$ & $48.055$    & $48.800^{\rm d}$, $51.698^{\rm e}$,
                                  $56.451^{\rm b}$ $49.166^{\rm f}$ &
                                                      $49.11(6)^{\rm h}$   \\
               \\
$^{137}$Ba$^+$
      & $6p_{3/2}$& $98.954$    & $92.275^{\rm i}$&$92.5(0.2)^{\rm j}$     \\
      &$5d_{3/2}$ & $45.765$    & $51.32^{\rm b}$, $47.3^{\rm k}$,
                                  $46.82^{\rm i}$ &$44.541(17)^{\rm k}$    \\
      &$5d_{5/2}$ & $62.685$    & $68.16^{\rm b}$, $63.2^{\rm k}$,
                                  $62.27^{\rm i}$ & $59.533(43)^{\rm k}$,
                                                    $60.7(10)^{\rm l}$,
                                                    $62.5(40)^{\rm m}$     \\
               \\
$^{173}$Yb$^+$
      &$6p_{3/2}$ & $1839.779$  & $1780^{\rm a}$    & $1460(50)^{\rm n}$   \\
      &$5d_{3/2}$ & $902.301$   & $951.4^{\rm b}$   & $-$                  \\
      &$5d_{5/2}$ & $1165.046$  & $1190.4^{\rm b}$  & $-$                  \\
\end{tabular}
\end{ruledtabular}
\begin{tabbing}
$^{\rm a}$ Reference\cite{Martensson-94}. \;\;\;\=  
                          $^{\rm b}$ Reference\cite{Itano-06}.      \\
$^{\rm c}$ Reference\cite{Krebs-55}.            \>
                          $^{\rm d}$ Reference\cite{Martensson-02}. \\
$^{\rm e}$ Reference\cite{Yu-04}.               \>
                          $^{\rm f}$ Reference\cite{Sahoo-07}.      \\
$^{\rm g}$ Reference\cite{Buchinger-90}.        \>
                          $^{\rm h}$ Reference\cite{Barwood-03}.    \\
$^{\rm i}$ Reference\cite{Sahoo-06}.            \>
                          $^{\rm j}$ Reference\cite{Villemoes-93}.  \\
$^{\rm k}$ Reference\cite{Silverans-86}.        \>
                          $^{\rm l}$ Reference\cite{Silverans-80}.  \\
$^{\rm m}$ Reference\cite{Hove}.                \>
                          $^{\rm n}$ Reference\cite{Berends-92}.    \\
\end{tabbing}
\end{table*}
Our result of $6s_{1/2}$ is slightly higher than both the theoretical and 
experimental values. However, for $6p_{1/2}$ our result is lower than the 
previous theoretical result of Martensson \cite{Martensson-94} and closer to 
the experimental data \cite{Martensson-94}. Similarly, our result of $6p_{3/2}$
is lower than the value of Martensson \cite{Martensson-94}. Although, the 
many-body methods employed in the two calculations are the same, HFS constants 
of $p$ in our results are lower.  Coming to the $5d$ states, the previous 
calculations of Itano is based on multiconfiguration Dirac-Fock (MCDF)
\cite{Parpia-96}. The $5d_{3/2}$ results are close to our value, however,
for the $5d_{5/2}$ state our results are much larger. At this stage it is 
difficult to pinpoint the reason for the large discrepancy between the 
two results. One observation from the component wise contribution in 
Table. \ref{tab-hfs-comp} is the large cancellation between the Dirac-Fock
(leading order) and the next to leading order 
$(S^\dagger \tilde H_{\rm hfs} + {\rm c.c.}) $. Similar pattern is observed in 
the HFS constant of $nd_{5/2}$ state of all the alkaline-Earth metal ions 
reported in our previous work \cite{Mani-10}. A comparison shows the 
cancellation is larger in Yb$^+$. Another notable difference in Yb$^+$ is,
the $(S_2^\dagger\tilde H S_1 + {\rm c.c.})$ is large and almost cancels with
$S_2^\dagger\tilde HS_2$.


\subsubsection{Hyperfine structure constant $b$ of $p_{3/2}$ states}

Leading terms, listed in Table. \ref{tab-hfs-comp}, are the Dirac-Fock and  
$(S^\dagger\tilde H_{\rm hfs} + {\rm c.c.})$. The later subsumes the 
core-polarization effects. For all the ions, Sr$^+$, Ba$^+$ and YB$^+$, the 
contributions from these two terms are almost equal. This is a significant 
deviation from the observed pattern in the magnetic dipole HFS constant 
\cite{Mani-10}, which noticeable for Yb$^+$ in Table. \ref{tab-hfs-comp}. Among
all the theoretical calculations our results for Sr$^+$ is in better agreement 
with the experimental data. For Ba$^+$, no previous theoretical works and 
experimental data are available. Ours is the first study on the 
electric quadrupole HFS constant of the $6p_{3/2}$ state. Our result of 
Yb$^+$ is higher than the previous theoretical results of Martensson-Pendrill
and collaborators \cite{Martensson-94} as well as the experimental results
of Berends and Maleki \cite{Berends-92}.

\begin{table*}[t]
\caption{Magnetic dipole and electric quadrupole HFS constants 
         contributions from different terms.}
\label{tab-hfs-comp}
\begin{ruledtabular}
\begin{tabular}{cccccccccc}
Ion& state &\multicolumn{6}{c}{Coupled-cluster terms}                   \\
\hline                                                                  \\
   &       & DF & $\tilde H_{\rm hfs}$-DF & $S^\dagger\tilde H_{\rm hfs}$
           & $S^\dagger_2\tilde H_{\rm hfs} S_1$
           & $S^\dagger_1\tilde H_{\rm hfs} S_1$ 
           & $S^\dagger_2\tilde H_{\rm hfs} S_2$ & Other terms & Norm   \\
   &       &    &               & $+ c.c.$ & $+ c.c.$ & & & &           \\
 \hline                                                                 \\
                        \\
   &       &    &   & HFS constant $A$&        &     &   & &            \\
$^{173}$Yb$^+$
       &$6s_{1/2}$&$-2582.096$&$130.175$&$-998.855$&$-31.463$&$-48.566$
                                      &$-60.815$&$14.667$  &$1.013$     \\
       &$6p_{1/2}$&$-408.696$&$15.693$&$-197.323$&$-7.054$ &$-10.972$
                                      &$-6.151$  &$-4.378$ &$1.011$     \\
       &$6p_{3/2}$&$-48.278$ &$1.798$ &$-32.418$ &$-1.817$ &$-1.234$
                                      &$-7.445$  &$-0.591$ & $1.011$    \\
       &$5d_{3/2}$&$-75.876$ &$0.192$ &$-19.903$ &$-0.870$ &$-1.289$
                                      &$-8.127$  &$-0.122$ &$1.015$     \\
       &$5d_{5/2}$&$-28.899$ &$-0.536$&$52.927$  &$4.081$  &$-0.424$
                                      &$-4.903$  &$0.082$  &$1.011$     \\
                        \\
   &       &    &   & HFS constant $B/Q$&        &     &   &  &         \\
$^{87}$Sr$^+$
       &$6p_{3/2}$&$166.993$&$-2.509$&$109.511$&$4.024$ &$2.565$
                                     &$7.769$  &$-2.302$&$1.001$        \\
       &$5d_{3/2}$&$80.939$ &$7.879$ &$26.888$ &$-.573$ &$1.073$
                                     &$-3.540$ &$-0.791$&$1.005$        \\
       &$5d_{5/2}$&$110.863$&$15.702$&$34.180$ &$-.848$ &$1.377$
                                     &$-1.911$ &$-1.081$&$1.005$        \\
                        \\
$^{137}$Ba$^+$
       &$6p_{3/2}$&$229.303$&$-5.962$&$170.985$ &$7.425$&$5.364$
                                      &$-1.153$  &$-3.097$ &$1.002$     \\
       &$5d_{3/2}$&$135.098$&$12.635$ &$46.886$ &$.211$ &$1.305$
                                      &$-7.849$  &$-1.213$ &$1.006$     \\
       &$5d_{5/2}$&$172.975$&$25.621$ &$63.015$ &$.333$ &$1.627$
                                      &$-5.898$ &$-1.508$&$1.005$       \\
                        \\
$^{173}$Yb$^+$
       &$6p_{3/2}$&$372.894$&$-16.855$&$274.025$ &$11.547$&$9.994$
                                      &$-0.212$  &$6.441$ &$1.001$      \\
       &$5d_{3/2}$&$199.032$&$11.143$ &$112.382$ &$1.292$ &$3.058$
                                      &$-4.128$  &$0.975$ &$1.005$      \\
       &$5d_{5/2}$&$234.438$&$21.220$ &$152.106$ &$2.217$ &$3.080$
                                      &$4.774$   &$-0.041$&$1.004$      \\
\end{tabular}
\end{ruledtabular}
\end{table*}


\subsubsection{Hyperfine structure constant $B$ of $d$ states}

 The HFS constant $B_d$ of Sr$^+$ has been studied in several theoretical
works. Our results of $B$ are systematically lower than the other theoretical 
values, which is evident from Table. \ref{tab-hfs}. The previous
calculations of Martensson \cite{Martensson-02} used the same many-body
method like ours, relativistic coupled-cluster, but a different type of 
single particle basis set. The calculations of Sahoo \cite{Sahoo-07}, in terms 
of theoretical methods, is closest to ours. They have used the relativistic 
coupled-cluster and gaussian type basis set like we have done. However, Sahoo 
calculated only for the $4d_{5/2}$ state and his result is closest to the 
experimental data. Ours on the other hand is $\approx 2.3\% $ lower than his 
result. Similarly, for the same reasons for the HFS constants $B$ of Ba$^+$, 
the previous calculations of Sahoo \cite{Sahoo-06} is closest to ours. However,
our result of $5d_{3/2}$  is closer to the experimental data. For $5d_{5/2}$ 
state our value is $\approx 0.6\%$ lower than the theoretical value of Sahoo 
and $\approx 5.3\% $ higher than the experimental value. Considering that for 
the Sr$^+$ and Ba$^+$ calculations, the many-body method and type of single 
particle basis we have used are the same as in ref. \cite{Sahoo-07} and 
\cite{Sahoo-06}, respectively.  The difference in the results could be on 
account of minor differences in the  exponents used in the basis set 
generation or the truncation of the coupled-cluster properties expression. 
There are striking changes, when compared with the $p_{3/2}$, in the component 
wise contribution. Dirac-Fock contribution in both, Sr$^+$ and Ba$^+$, are
approximately three times larger than 
$(S^\dagger\tilde H_{\rm hfs} + {\rm c.c.}$.
In addition, the contribution from the $\tilde H_{\rm hfs}-DF $, which
essentially arises from the closed-shell part, is relatively large. This could
be due to the diffuse electron density of the $d$ orbitals and hence 
stronger interaction with the core electrons.

 Unlike the other two ions, Yb$^+$ has not been studied in fine detail. The 
previous theoretical work of Itano \cite{Itano-06} is based on the 
MCDF method. And there are no experimental data
available for the $^{173}$Yb$^+$ isotope. Our results are lower but close to
the values from Itano \cite{Itano-06}. A closer inspection of the results from
Itano's calculations for the other ions (Sr$^+$ and Ba$^+$) reveals that, 
his results are consistently higher than the other theoretical and experimental
data. One possible reason could be the contracted nature of the virtual 
orbitals, referred as correlation orbitals, in MCDF calculations. Hence we can 
expect a similar trend in Yb$^+$ as well and it is possible that our results 
are closer to the actual values. Compared to Sr$^+$ and Ba$^+$, there is one 
remarkable change in the component wise contribution. There is large  
contribution from $(S^\dagger\tilde H_{\rm hfs} + {\rm c.c.})$, which implies 
that there core-polarization effect is very important. It is on par with the 
Dirac-Fock term.


\subsubsection{Two-valence}

There are few theoretical and experimental work on the HFS constants of the 
neutral alkaline-Earth metal atoms and Yb. However, the importantance of such
investigations are likely grow in the near future as these, in particular
Sr and Yb, are candidates of precision experiments and have been cooled to 
quantum degeneracy. In our work we make an effort to understand the 
systematics to initiate a deeper analysis on the role of the electron 
correlation effects to properties like hyperfine. The previous theoretical
calculations of $^{137}$Ba \cite{Kozlov-99} and $^{173}$Yb \cite{Porsev-99} 
are based on the CI-MBPT method and basis with different central potentials. 
\begin{table*}[t]
\caption{Magnetic dipole HFS constant for the atomic systems $^{87}$Sr, 
         $^{137}$Ba, and $^{173}$Yb, using relativistic coupled-cluster
         theory. All values are in atomic units.}
\label{tab-hfs-2v}
\begin{ruledtabular}
\begin{tabular}{cccccccc}
 State& \multicolumn{5}{c}{Coupled-cluster terms}& Other work & Exp result \\
  \hline                                                                   \\
      & DF & $\tilde H_{\rm hfs}$-DF & One-body $\tilde H_{\rm hfs}$
      & Two-body $\tilde H_{\rm hfs}$& Total value &    &                  \\
  \hline                                                                   \\
Atomic $^{87}$Sr$;\;$[Kr]$5s^2$&  & &   &   &    &    &     \\
$5s5p\;^3P_1$ & $-178.983$ & $-0.120$   & $-49.121$ & $0.002$ 
                           & $-228.222$ & $-$        & $-260.765(1)^{\rm j}$ \\
$5s5p\;^3P_2$ & $-200.670$ & $0.106$    & $-47.045$ & $0.002$  
                           & $-247.607$ & $-$       & $-212.085(5)^{\rm j}$ \\
$5s4d\;^3D_1$ & $145.335$  & $0.098$    & $6.348$   & $0.001$ 
                           & $151.586$  & $-$       & $-$                   \\
$5s4d\;^3D_2$ & $-56.824$  & $0.076$    & $7.095$   & $0.001$ 
                           & $-49.654$  & $-$       & $-$                   \\
$5s4d\;^3D_3$ & $-133.930$ & $-0.040$   & $.194$    & $0.002$ 
                           & $-133.778$ & $-$       & $-$                   \\
$5s4d\;^1D_2$ & $17.441$   & $0.062$    & $9.643$   & $0.001$ 
                           & $27.145$   & $-$       & $-$                   \\
$5s4d\;^1P_1$ & $11.802$   & $-0.225$   & $4.366$   & $-0.002$ 
                           & $15.941$   & $-$       & $-$                   \\
                           \\
Atomic $^{137}$Ba$;\;$[Xe]$6s^2$  &        &   &         &   &   &   &      \\ 
$6s5d\;^3D_1$ & $-588.432$ & $0.344$    & $-19.420$ & $-0.004$ 
                           & $-607.512$ & $-547^{\rm a}$ &$-521^{\rm c}$    \\
$6s5d\;^3D_2$ & $397.451$  & $-0.588$   & $9.067$   & $-0.009$ 
                           & $405.921$  & $405^{\rm a}$ & $416^{\rm c}$     \\
$6s5d\;^3D_3$ & $543.921$  & $0.189$    & $4.843$   & $-0.008$ 
                           & $548.945$  & $443^{\rm a}$ &$457^{\rm c}$      \\
$6s5d\;^1D_2$ & $-148.545$ & $-0.459$   & $-54.967$ & $-0.006$ 
                           & $-203.977$ & $-102^{\rm a}$& $-82^{\rm d}$     \\
$6s6p\;^3P_1$ & $736.066$  & $-0.310$   & $221.943$ & $-0.004$ 
                           & $957.695$  &$1160^{\rm a}$ &$1151^{\rm e}$     \\ 
$6s6p\;^3P_2$ & $806.032$  & $-0.152$   & $204.186$ & $-0.010$ 
                           & $1010.056$ & $845^{\rm a}$ & $-$               \\
$6s6p\;^1P_1$ & $-181.658$ & $-0.074$   & $-38.353$ & $0.009$ 
                           & $-220.094$ & $-107^{\rm a}$ & $-109^{\rm f}$   \\ 
                           \\
Atomic $^{173}$Yb$;\;$[Xe]$4f^{14}6s^2$&   &   &  &  & &   &                \\  
$6s6p\;^3P_1$ & $-708.922$ & $-0.223$  & $-197.665$& $0.004$
              & $-906.806$ & $-1094^{\rm b}$&$-1094.2(6)^{\rm g}$  \\
$6s6p\;^3P_2$ & $-681.732$ & $-0.292$  & $-181.832$& $0.003$
              & $-863.853$ & $-745^{\rm b}$ &$-738^{\rm h}$        \\
$6s5d\;^3D_1$ & $550.679$  & $0.035$&  $ 64.941$   & $0.002$
              & $615.657$  & $596^{\rm b}$  &$563(1)^{\rm i}$      \\
$6s5d\;^3D_2$ & $-456.152$ & $-0.064$  & $-40.287$ & $0.003$
              & $-496.500$ & $-351^{\rm b}$ &$-362(2)^{\rm i}$     \\ 
$6s5d\;^3D_3$ & $-454.431$ & $-0.027$  & $-22.482$ & $0.002$ 
              & $-476.938$ & $-420^{\rm b}$ & $-430(1)^{\rm i}$    \\ 
$6s6p\;^1P_1$ & $239.530$  & $0.499$   & $65.911$  & $-0.002$ 
              & $305.938$  & $191^{\rm b}$  & $60^{\rm h}$         \\  
$6s5d\;^1D_2$ & $197.218$  & $-0.011$  & $71.840$  & $0.002$ 
              & $269.049$  & $131^{\rm b}$ &$100(18)^{\rm i}$      \\ 
\end{tabular}
\end{ruledtabular}
\begin{tabbing}
$^{\rm a}$ Reference\cite{Kozlov-99}. \;\;\;\=  
                          $^{\rm b}$ Reference\cite{Porsev-99}.      \\
$^{\rm c}$ Reference\cite{Gustavsson-79}.   \> 
                          $^{\rm d}$ Reference\cite{Schmelling-74}.  \\
$^{\rm e}$ Reference\cite{Putliz-63}.       \>
                          $^{\rm f}$ Reference\cite{Kluge-74}.       \\
$^{\rm g}$ Reference\cite{Jin-91}.          \>
                          $^{\rm h}$ Reference\cite{Budick-69}.      \\
$^{\rm i}$ Reference\cite{Topper-97}.       \>
                          $^{\rm j}$ Reference\cite{Stephan-77}.
\end{tabbing}
\end{table*}
For this reason it is non trivial to comment on the role of the correlation
effects in a precise manner through a comparative study. The results from 
our calculations, along with the leading order contributions, are listed
in Table. \ref{tab-hfs-2v}.  From the table it is clear that, in most of the 
cases our theoretical results are not in very good agreement with the 
experimental data.  Origin of the discrepancy could be the nature of the 
single particle basis we have used, the $V^{N-2}$ orbitals. On account of
the doubly ionized charged state of the core, the orbitals are highly 
contracted and interacts rather strongly with the nucleus. Such orbitals are
suitable for properties calculations of singly ionized states but not 
ideal for the neutral atoms.

 A very important aspect of our present work is the observed trend in the
contributions from various terms.  It is evident from Table. \ref{tab-hfs-2v}
the DF contribution is significantly dominant, it is far larger than the 
next to leading order contribution from what we refer as the {\em one-body 
terms}. The details of the {\em one-body terms } are discussed in Section.
\ref{hfs-one-body}. To quantify the relative contributions, define
\begin{equation}
    \varrho = \frac{\text{one-body terms}}{\text{DF}}.
\end{equation}
Essentially the ratio between the leading and next to leading order 
contributions.  The dominance of DF is particularly true in the case of the 
$ns(n-1)d\;^3D_J $ states, among these states highest $\varrho$ is 
$\approx 0.12$ ( $ 5s4d\;^3D_2 $ state of $^{87}$Sr). For the $nsnp\; ^3P_J$ 
states, the contribution from the {\em one-body terms} is small but not 
negligible. Largest and smallest value of $\varrho$ for these states are 
$\approx 0.3$  for the $6s6p\;^3P_1$ state of Ba and  $\approx 0.23 $ for 
the $5s5p\; ^3P_2 $ state of Sr, respectively. Other states have $\varrho$ 
close to 0.25.

For the singlet states $nsnp\; ^1P_1$ and $ns(n-1)d\; ^1D_2$, the deviations
from the experimental data are very large. A similar trend was also observed in
the case of the excitation energy of these states as well. 


\subsection{E1 transition amplitude}

We calculate the reduced matrix element of the dipole operator $\mathbf{D}$ 
from the expression given in Eq. (\ref{d_if}). Once again, like in the HFS 
constants, we calculate the reduced matrix elements of the Sr$^+$, Ba$^+$
and Yb$^+$ ions from the intermediate one-valence wave functions. In present
work, we do not attempt to quantify the error or accuracy of the results. 
This is a work in progress and we shall report in our future publications with
a careful examination of the different types of basis functions. And,
calculate the dipole matrix elements in different gauges.


\subsubsection{One-valence}

 Results from our calculations are listed in Table. \ref{tab-dip} and 
component wise contributions are given in Table. \ref{tab-dip-comp}.
One of the early works on the dipole matrix elements of singly ionized 
alkaline-Earth metal ions is by Guet and Johnson \cite{Guet-91}. The many-body
method they used is MBPT and RPA, and numerical basis set. At the DF level
the values of Guet and Johnson \cite{Guet-91} are in good agreement, 
for both Sr$^+ $ and Ba$^+$, with our results, this is evident from the values 
listed in Table. \ref{tab-dip-comp}. Their work is the only one in the 
literature on the electric dipole matrix elements of Sr$^+$ and our results
are higher. The difference could be largely attributed to the higher order
core-polarization effects associated with the random-phase approximation (RPA).
The RPA effects are incorporated in the coupled-cluster but not  to higher 
order as in an iterative RPA calculations.

\begin{table}[t]
\caption{Magnitude of the electric dipole transition amplitude 
         for $^{87}$Sr$^+$, $^{137}$Ba$^+$, and $^{173}$Yb$^+$ ions.}
\label{tab-dip}
\begin{ruledtabular}
\begin{tabular}{cccc}
                                                                     \\
Ion & Transition                  & This work         & Other works  \\
\hline
                 \\
$^{87}$Sr$^+$&
  $5p_{1/2}\longrightarrow 5s_{1/2}$&$3.2180$&$3.060$\footnotemark[1] \\
& $5p_{3/2}\longrightarrow 5s_{1/2}$&$4.9223$&$4.325$\footnotemark[1] \\
& $5p_{1/2}\longrightarrow 4d_{3/2}$&$3.4315$&$3.052$\footnotemark[1] \\
& $5p_{3/2}\longrightarrow 4d_{3/2}$&$1.4217$&$1.355$\footnotemark[1] \\
& $5p_{3/2}\longrightarrow 4d_{5/2}$&$4.5942$&$4.109$\footnotemark[1] \\
                 \\
$^{137}$Ba$^+$&
  $6p_{1/2}\longrightarrow 6s_{1/2}$&$3.1974$ &$3.300$\footnotemark[1],
                                              $3.36(1)$\footnotemark[2],
                                              $3.272$\footnotemark[3] \\
& $6p_{3/2}\longrightarrow 6s_{1/2}$&$5.0330$ &$4.658$\footnotemark[1],
                                              $4.73(3)$\footnotemark[2],
                                              $4.614$\footnotemark[3] \\
& $6p_{1/2}\longrightarrow 5d_{3/2}$&$3.0898$ &$3.009$\footnotemark[1],
                                              $3.11(3)$\footnotemark[2],
                                              $3.008$\footnotemark[3] \\
& $6p_{3/2}\longrightarrow 5d_{3/2}$&$1.2448$ &$1.312$\footnotemark[1],
                                              $1.34(2)$\footnotemark[2],
                                              $1.313$\footnotemark[3] \\
& $6p_{3/2}\longrightarrow 5d_{5/2}$&$4.1347$ &$4.057$\footnotemark[1],
                                              $4.02(7)$\footnotemark[2],
                                              $4.054$\footnotemark[3] \\
                 \\
$^{173}$Yb$^+$&
  $6p_{1/2}\longrightarrow 6s_{1/2}$&$2.9069$&$2.731$\footnotemark[4] \\
& $6p_{3/2}\longrightarrow 6s_{1/2}$&$4.5256$&$3.845$\footnotemark[4] \\
& $6p_{1/2}\longrightarrow 5d_{3/2}$&$3.6317$&$3.782$\footnotemark[4] \\
& $6p_{3/2}\longrightarrow 5d_{3/2}$&$1.4918$&$1.546$\footnotemark[4] \\
& $6p_{3/2}\longrightarrow 5d_{5/2}$&$4.8500$&$4.769$\footnotemark[4] \\
\end{tabular}
\end{ruledtabular}
\footnotetext[1]{Reference\cite{Guet-91}.}
\footnotetext[2]{Reference\cite{Sahoo-09}.}
\footnotetext[3]{Reference\cite{Dzuba-01}.}
\footnotetext[4]{Reference\cite{Safronova-09}.}
\end{table}
 For Ba$^+$ there are several theoretical calculations of the electric 
dipole matrix elements. A careful study on the electric dipole transition
is desirable as it a promising candidate for a novel parity non-conservation
experiment \cite{Fortson-93}. In terms of the many-body method and single 
particle basis set, the calculations of Sahoo and collaborators 
\cite{Sahoo-09} are closest to our approach. They estimate the upper bound
on the error in the reduced dipole matrix element as 1.7\%, which implies
that our results have errors larger than this.

For Yb$^+$, the work of Safronova and Safronova \cite{Safronova-09} is the only
previous study on the electric dipole matrix elements. Their calculations
are based on the third order relativistic MBPT and the excellent matching
between the length and velocity gauge results indicates the results are quite
accurate. Our results are close to their results, however, at this stage
we do not attempt to estimate the accuracy of our results.

\begin{table*}[h]
\caption{The electric dipole transition amplitude, contributions from 
         different terms in the coupled-cluster theory.}
\label{tab-dip-comp}
\begin{ruledtabular}
\begin{tabular}{cccccccccc}
Ion &Transition&\multicolumn{6}{c}{Coupled-cluster terms}                  \\
\hline                                                                     \\
    &     & DF & $\tilde D$-DF & $S^\dagger\tilde D$
          & $S^\dagger_2\tilde D S_1$
          & $S^\dagger_1\tilde D S_1$ 
          & $S^\dagger_2\tilde D S_2$ & Other terms & Norm                 \\
    &     &    &  & $+ c.c.$ & $+ c.c.$ & & &                              \\
 \hline                                                                    \\
                        \\
$^{87}$Sr$^+$&
 $5p_{1/2}\longrightarrow 5s_{1/2}$&$3.4869$&$0.0008$&$-0.2715$&$-0.0043$
                                   &$0.0129$ &$0.0233$&$-0.0004$&$0.9909$  \\
& $5p_{3/2}\longrightarrow 5s_{1/2}$&$4.9246$&$0.0019$&$-0.0072$&$-0.0003.1$
                                    &$0.0187$&$0.0034$&$-0.0047$&$0.9902$  \\
& $5p_{1/2}\longrightarrow 4d_{3/2}$&$3.7226$&$0.0024$&$0.2902$&$-0.0062$
                                   &$0.0178$&$0.0234$&$0.0031$&$0.9889$    \\
& $5p_{3/2}\longrightarrow 4d_{3/2}$&$1.6543$&$0.0001$ &$-0.2332$&$-0.0028$
                                    &$0.0080$&$0.0122$&$0.0002$&$0.9882$   \\
& $5p_{3/2}\longrightarrow 4d_{5/2}$&$-4.9937$&$-0.0005$&$0.3967$&$0.0085$
                                    &$-0.0238$&$-0.0334$&$-0.0006$&$0.9887$\\
                        \\
$^{137}$Ba$^+$&
  $6p_{1/2}\longrightarrow 6s_{1/2}$&$3.8911$&$0.0019$&$-0.7618$&$-0.0097$
                                    &$0.0442$&$0.0715$&$-0.0009$&$0.9880$  \\
& $6p_{3/2}\longrightarrow 6s_{1/2}$&$-5.4778$&$-0.0046$&$0.5275$&$0.0134$
                                    &$-0.0609$&$-0.0973$&$0.0014$&$0.9872$ \\
& $6p_{1/2}\longrightarrow 5d_{3/2}$&$-3.7450$&$-0.0080$&$0.7220$&$-0.0001$
                                    &$-0.0392$&$-0.0685$&$0.0008$&$0.9846$ \\
& $6p_{3/2}\longrightarrow 5d_{3/2}$&$1.6352$&$0.0119$&$-0.4240$&$0.0005$
                                    &$0.0161$&$0.0363$&$0.0000$&$0.9838$   \\
& $6p_{3/2}\longrightarrow 5d_{5/2}$&$5.0005$&$0.0102$&$-0.9544$&$0.0011$
                                    &$0.0485$&$0.0930$&$-0.0001$&$0.9847$  \\
                        \\
$^{173}$Yb$^+$&
  $6p_{1/2}\longrightarrow 6s_{1/2}$&$3.2422$&$0.0011$&$-0.3387$&$-0.0071$
                                    &$0.0181$&$0.0247$&$0.0043$&$0.9872$   \\
& $6p_{3/2}\longrightarrow 6s_{1/2}$&$-4.5426$&$-0.0032$&$0.0282$&$-0.0001$
                                    &$-0.0231$&$-0.0430$&$-0.0021$&$0.9868$\\
& $6p_{1/2}\longrightarrow 5d_{3/2}$&$-3.8611$&$-0.0024$&$0.2336$&$0.0095$
                                    &$-0.0286$&$-0.0366$&$0.0055$&$0.9869$ \\
& $6p_{3/2}\longrightarrow 5d_{3/2}$&$.6970$&$0.0002$&$-0.2551$&$-0.0039$
                                    &$0.0114$&$0.0165$&$-0.0022$&$0.9865$  \\
& $6p_{3/2}\longrightarrow 5d_{5/2}$&$-5.2002$&$0.0008$&$0.3448$&$0.0117$
                                    &$-0.0325$&$-0.0443$&$0.0113$&$0.9881$ \\
\end{tabular}
\end{ruledtabular}
\end{table*}


\subsubsection{Two-valence}
 
Our results of the dipole matrix elements of Sr, Ba and Yb are listed in the
tables Table. \ref{tab-dip-sr}-\ref{tab-dip-yb}. There are very few 
theoretical studies on the dipole matrix elements of Sr and those are not 
in very good agreement with ours. In the case of Ba, the previous theoretical
calculations were done by Dzuba and Ginges \cite{Dzuba-06}. Our results,
listed in Table. \ref{tab-dip-ba} are good agreement with ref. \cite{Dzuba-06}
for the $\langle 6s6p\; ^3D_1||D|| 6s^2\; ^1S_0 \rangle$ and 
$ \langle 6s6p\; ^3D_J||D||6s5d\; ^3D_J \rangle$. However, there
are large deviations for the matrix elements involving the $6s6p\; ^1P_1$ and 
$6s5d\;^1D_2$ states. For Yb, the $\langle 6s6p\;^1P_1||D||6s5d\;^3D_2\rangle$ 
is significantly different from the previous results. However, there is a 
large difference between the previous results of Porsev and collaborators
\cite{Porsev-99b}, and Migdalek and Baylis \cite{Migdalek-91} as well. Large
relative deviations, compared to results of ref. \cite{Porsev-99b}, are also 
observed for the $\langle 6s6p\;^3P_1||D||6s5d\;^1D_2\rangle$ and 
$\langle 6s6p\;^3P_2||D||6s5d\;^1D_2\rangle$.

\begin{table*}[th]
\caption{E1 transition amplitudes for the atomic system $^{87}$Sr, 
         using relativistic coupled-cluster theory. All values are
         in atomic units.}
\label{tab-dip-sr}
\begin{ruledtabular}
\begin{tabular}{ccccccc}
Transition & \multicolumn{5}{c}{Coupled-cluster terms}& Other work      \\
  \hline                                                                \\
     & DF & $\tilde H_{\rm hfs}$-DF & One-body $\tilde H_{\rm hfs}$
     & Two-body $\tilde H_{\rm hfs}$& Total value &                     \\
  \hline                                                                \\
$^3P_1\longrightarrow\;^1S_0$&$-0.3759$  &$-0.0001$&$0.8257$  &$0.00001$
                             &$0.4497$ &$0.16$\footnotemark[1],
                                        $0.162$\footnotemark[2]         \\
$^1P_1\longrightarrow\;^1S_0$&$-4.2442$  &$0.0000$ &$0.5283$  &$-0.00002$
                             &$-3.7159$&$5.28$\footnotemark[1],
                                        $5.238$\footnotemark[2],
                                        $1.9539$\footnotemark[3]        \\
$^3P_0\longrightarrow\;^3D_1$&$2.6323$  &$-0.0002$&$-0.3131$ &$0.00000$
                             &$2.3190$ &$$                              \\
$^3P_1\longrightarrow\;^3D_1$&$2.2652$  &$0.0013$ &$0.1116$  &$0.00000$
                             &$2.3781$ &$$                              \\
$^3P_2\longrightarrow\;^3D_1$&$0.5849$  &$0.0009$ &$0.3772$  &$-0.00001$
                             &$0.9630$ &$$                              \\
$^1P_1\longrightarrow\;^3D_1$&$0.2255$  &$-0.0005$&$-0.2496$ &$-0.00001$
                             &$-0.0247$&$$                              \\
$^3P_1\longrightarrow\;^3D_2$&$3.9538$  &$0.0012$ &$-0.5437$ &$-0.00001$
                             &$3.4114$ &$$                              \\
$^3P_2\longrightarrow\;^3D_2$&$2.2646$  &$0.0006$ &$0.0700$  &$0.00001$
                             &$2.3352$ &$$                              \\
$^1P_1\longrightarrow\;^3D_2$&$0.7531$  &$-0.0004$&$-0.1054$ &$0.00001$
                             &$0.6473$ &$$                              \\
$^3P_2\longrightarrow\;^3D_3$&$-5.3938$  &$-0.0001$&$0.3111$ &$0.00001$
                             &$-5.0828$ &$$                             \\
$^3P_1\longrightarrow\;^1D_2$&$-0.8822$  &$0.0012$ &$0.0387$ &$0.00000$
                             &$-0.8423$&$0.19$\footnotemark[1]          \\
$^3P_2\longrightarrow\;^1D_2$&$-0.2854$  &$0.0002$ &$-0.3597$&$-0.00001$
                             &$-0.6448$&$0.10$\footnotemark[1]          \\
$^1P_1\longrightarrow\;^1D_2$&$-4.5484$  &$-0.0008$&$0.3860$ &$-0.00001$
                             &$-4.1632$&$1.92$\footnotemark[1]          \\
\end{tabular}
\end{ruledtabular}
\footnotetext[1]{Reference\cite{Porsev-01}.}
\footnotetext[2]{Reference\cite{Savukov-02}.}
\footnotetext[3]{Reference\cite{Vaeck-88}.}
\end{table*}
\begin{table*}[th]
\caption{E1 transition amplitudes for the atomic system $^{137}$Ba, 
         using relativistic coupled-cluster theory. All values are
         in atomic units.}
\label{tab-dip-ba}
\begin{ruledtabular}
\begin{tabular}{ccccccc}
Transition & \multicolumn{5}{c}{Coupled-cluster terms}& Other work         \\
  \hline                                                                   \\
     & DF & $\tilde H_{\rm hfs}$-DF & One-body $\tilde H_{\rm hfs}$
     & Two-body $\tilde H_{\rm hfs}$& Total value &                        \\
  \hline                                                                   \\
$^3P_1\longrightarrow\;^1S_0$&$0.3888$ &$-0.0003$&$-0.8090$  &$0.00000$
                                       &$0.4205$&$0.4537$\footnotemark[1]  \\
$^1P_1\longrightarrow\;^1S_0$&$-4.6768$&$-0.0002$&$0.6730$  &$0.00000$  
                                       &$4.0040$ &$5.236$\footnotemark[1]  \\
$^3P_0\longrightarrow\;^3D_1$&$2.6203$ &$0.0004$ &$-0.2585$  &$0.00000$  
                                       &$2.3622$ &$2.3121$\footnotemark[1] \\
$^3P_1\longrightarrow\;^3D_1$&$2.2405$ &$-0.0023$&$-0.0195$  &$0.00000$  
                                       &$2.2187$ &$2.0108$\footnotemark[1] \\
$^3P_2\longrightarrow\;^3D_1$&$-0.5715$&$0.0019$ &$-0.4782$  &$0.00000$  
                                       &$1.0478$ &$0.5275$\footnotemark[1] \\
$^1P_1\longrightarrow\;^3D_1$&$-0.3364$&$-0.0002$&$0.3546$  &$0.00000$  
                                       &$0.0180$ &$0.1047$\footnotemark[1] \\
$^3P_1\longrightarrow\;^3D_2$&$3.8886$ &$-0.0022$&$-0.1585$  &$0.00001$  
                                       &$3.7279$ &$3.4425$\footnotemark[1] \\
$^3P_2\longrightarrow\;^3D_2$&$-2.2265$&$0.0013$ &$0.1007$  &$-0.00001$  
                                       &$2.1245$ &$2.024$\footnotemark[1]  \\
$^1P_1\longrightarrow\;^3D_2$&$-0.3874$&$-0.0005$&$0.0833$  &$-0.00000$  
                                       &$0.3046$ &$0.4827$\footnotemark[1] \\
$^3P_2\longrightarrow\;^3D_3$&$5.3410$ &$-0.0004$&$-0.3409$  &$-0.00001$  
                                       &$4.9997$ &$4.777$\footnotemark[1]  \\
$^3P_1\longrightarrow\;^1D_2$&$-1.1039$&$-0.0018$&$0.1178$  &$0.00000$  
                                       &$0.9879$ &$0.1610$\footnotemark[1] \\
$^3P_2\longrightarrow\;^1D_2$&$0.4458$ &$0.0005$ &$0.5219$  &$0.00001$  
                                       &$0.9682$ &$0.1573$\footnotemark[1] \\
$^1P_1\longrightarrow\;^1D_2$&$4.4933$ &$-0.0011$&$-0.0773$  &$0.00000$  
                                       &$4.4149$ &$1.047$\footnotemark[1]  \\
\end{tabular}
\end{ruledtabular}
\footnotetext[1]{Reference\cite{Dzuba-06}.}
\end{table*}

\begin{table*}[th]
\caption{E1 transition amplitudes for the atomic system $^{173}$Yb, 
         using relativistic coupled-cluster theory. All values are
         in atomic units.}
\label{tab-dip-yb}
\begin{ruledtabular}
\begin{tabular}{ccccccc}
Transition & \multicolumn{5}{c}{Coupled-cluster terms}& Other work         \\
  \hline                                                                   \\
     & DF & $\tilde H_{\rm hfs}$-DF & One-body $\tilde H_{\rm hfs}$
     & Two-body $\tilde H_{\rm hfs}$& Total value &                        \\
  \hline                                                                   \\
$^3P_1\longrightarrow\;^1S_0$&$0.1445$&$-0.0003$ &$-0.5320$&$0.00000$ 
                                      &$-0.3878$&$0.54(8)$\footnotemark[1],
                                                 $0.44$\footnotemark[2],
                                                 $0.587$\footnotemark[3]    \\
$^1P_1\longrightarrow\;^1S_0$&$-3.8641$&$-0.0001$&$0.5999$&$-0.00001$ 
                                      &$-3.2643$&$4.40(80)$\footnotemark[1],
                                                 $4.44$\footnotemark[2],
                                                 $4.89$\footnotemark[4],
                                                 $4.825$\footnotemark[3]    \\
$^3P_0\longrightarrow\;^3D_1$&$2.7296$&$0.0001$  &$-0.3209$&$0.00000$ 
                                      &$2.4088$ &$2.61(10)$\footnotemark[1],
                                                 $2.911$\footnotemark[3]\\
$^3P_1\longrightarrow\;^3D_1$&$2.3473$&$-0.0005$ &$0.1811 $&$0.00003$ 
                                      &$2.5279$ &$2.26(10)$\footnotemark[1]\\
$^3P_2\longrightarrow\;^3D_1$&$-0.5997$&$0.0000$ &$-0.2343$&$0.000002$ 
                                      &$-0.8340$&$0.60(12)$\footnotemark[1]\\
$^1P_1\longrightarrow\;^3D_1$&$-0.4503$&$-0.0002$&$0.1702$&$0.00000$ 
                                      &$-0.2803$&$0.27(10)$\footnotemark[1],
                                                 $0.24$\footnotemark[2]    \\
$^3P_1\longrightarrow\;^3D_2$&$3.9875$&$-0.0005$ &$-0.6480$&$0.00002$ 
                                      &$3.3390$ &$4.03(16)$\footnotemark[1]\\
$^3P_2\longrightarrow\;^3D_2$&$-2.2940$&$-0.0002$&$-0.1010$&$-0.00003$ 
                                      &$-2.3952$&$2.39(1)$\footnotemark[1] \\
$^1P_1\longrightarrow\;^3D_2$&$0.0716$&$-0.0003$ &$0.0660$&$0.00000$ 
                                      &$0.1373$ &$0.32(6)$\footnotemark[1],
                                                 $0.60$\footnotemark[2]    \\
$^3P_2\longrightarrow\;^3D_3$&$5.6130$&$0.0000$  &$-0.3215$&$-0.00001$ 
                                      &$5.2915$ &$6.12(30)$\footnotemark[1]\\
$^3P_1\longrightarrow\;^1D_2$&$-1.1920$&$-0.0006$&$0.0995$&$0.00000$ 
                                      &$-1.0931$&$0.54(10)$\footnotemark[1]\\
$^3P_2\longrightarrow\;^1D_2$&$0.5946$&$0.0002$  &$0.3601$&$0.00000$ 
                                      &$0.9549$ &$0.38(8)$\footnotemark[1] \\
$^1P_1\longrightarrow\;^1D_2$&$4.7006$&$-0.0002$ &$-0.5209$&$0.00000$ 
                                      &$4.1795$ &$3.60(70)$\footnotemark[1]\\
\end{tabular}
\end{ruledtabular}
\footnotetext[1]{Reference\cite{Porsev-99b}.}
\footnotetext[2]{Reference\cite{Migdalek-91}.}
\footnotetext[3]{Reference\cite{Dzuba-10}.}
\footnotetext[4]{Reference\cite{Kunisz-82}.}
\end{table*}


\section{Conclusions}

In this paper we describe in detail the Fock-space relativistic 
coupled-cluster method for the two-valence systems. It is based on an all 
particle treatment and we demonstrate the excitation energies of 
Ba and Yb are on par with those of the previous relativistic CC 
calculations. The key point is, we have implemented the Fock-space CCT with 
an incomplete but quasi-complete model space comprising of the $ns^2$, 
$ns(n-1)d$ and $nsnp$ configurations. This choice of model space is optimal
to study the low-lying states of the alkaline-Earth metal atoms and other 
two-valence atoms like Yb and Hg. Most importantly, with this model space
one may avoid divergences arising from the {\em intruder} states. This ought
to be highlighted as the literature on iterative studies of 
two-valence systems is replete with accounts of {\em intruder} state induced
divergences. We emphasize that there are few detailed relativistic many-body
calculations of two-valence excitation energies and even less on properties
calculations. Not surprisingly, among the published results there is a wide 
variation of the many-body methods and basis sets used in the studies. 
Considering the growing importance of alkaline-Earth metal atoms in precision 
experiments and possible applications, a detailed investigations on the 
two-valence systems is timely.

  We have also developed a method based on CCT to compute properties from the 
CC wave functions of two-valence systems. This is perhaps an initial step 
towards systematic investigation of the structure and properties of 
two-valence systems with CCT, which has not been attempted before. Based on our
scheme, the comptutational cost of two-valence CC calculations is marginally 
higher than the one-valence calculations. And the additional cost is in solving 
the $S^{(2)}$ cluster amplitude equations. Total number of which is far less 
than the closed-shell and one-valence cluster amplitudes $T$ and $S^{(1)}$, 
respectively. So in terms of computational implementations, there is no reason 
why two-valence CCT should not be the preferred method as in 
one-valence systems. The important and essential details of our schemes and 
implmentations are provided to highlight important physics issues in the 
two-valence Fock-space CCT. Breif descriptions of the method and extensive
references are provided to aid interested researchers to implement CCT of
two-valence systems.

 From the many-body theory perspective, CCT based structure and properties 
calculations is certainly an attractive choice. Prime reason being the 
topologically connected nature of the CC operators and exponential form of
the wave operator ensures the condition of size extensivity. A basic 
requirement of a legitimate many-body theory. Further more, the 
non-perturbative character of the CC wave operator makes it an ideal choice. 
It must be emphasized that, all of these considerations are at the
level of the many-body theory. However, the accuracy of the results also 
depends on other factors like single particle basis set considered. 

  In the method we have developed, the one-valence coupled-cluster wave 
functions occur as an intermediate step. Using this we calculate the electric 
quadrupole HFS constant $B$ of Sr$^+$, Ba$^+$ and Yb$^+$. 
The HFS constant $B$ of the $5p_{3/2}\;^2P_{3/2}$  and $5d_{3/2}\;^2D_{3/2}$ 
states of $^{87}$Sr$^+$ and $^{137}$Ba$^+$, respectively are closer to the 
experimental data than the other theoretical results. For all the ions studied,
the $np_{3/2}\; ^2P_{3/2}$ state has very large contributions from the 
core polarization effects, which is part of
$(S^{\dagger}\tilde H_{\rm hfs} + {\rm c.c.})$  in the CC properties 
calculations. Similaraly, for $5d_j\; ^2D_{j}$ states of Yb$^+$, there is a 
large contribution from the core-polarization effects. A careful
accounting of the core-polarization effects is crucial for all the ions to
obtain accurate values of $B$ and this is particularly true for Yb$^+$.
We also calculate the magnetic dipole HFS constant $A$ 
of Yb$^+$, except for the $5d_{5/2}$ state the results are in agreement 
with the other theoretical and experimental data. Similarly, we get reliable
results of the dipole matrix elements of these ions. 

 The results of the excitation energies of the two-valence sector calculations 
are in agreement with the previous CC results. For properties calculations
with CC wave functions, our results show deviations from the previous works. 
However, it must be mentioned that there have been very few attempts at 
theoretical properties calculations of two-valence systems. And the previous 
works are based on MCDF or using a collage of single particle wave functions. 
The later may require finer analysis for precision studies as the linked-cluster
theorem, which forms the basis of many-body theory, is based on a uniform
separation of the total Hamiltonian into zeroth order and perturbation. This is
not the case when orbitals with different central potentials are used in the 
calculations. With MCDF method, a large scale structure and properties 
calculations of neutral atoms  with high $Z$ is plagued with convergence 
issues. Among all the states, the singlet states $nsnp\; ^1P_1$ 
and $ns(n-1)d\; ^1D_2$ require further attention as the properties  involving
these state exhibit largest deviations from other theoretical results. A
similar pattern is observed in the other theoretical results as well.  
Based on our studies and careful analysis, the observed deviations of the
two-valence properties may be attributed to the basis, $V^{N-2}$ potential,
we have used. We expect calculations with $V^{N-1}$ potential basis
could improve the results. 

 In conclusion, relativistic Fock-space coupled-cluster theory has theoretical
and computational advantages for structure and properties calculations of 
two-valence systems. In this article we report the development of
an all particle two-valence relativistic Fock-space coupled-cluster theory and
have demonstrated a scheme for properties calculations with the CC wave 
functions.


\begin{acknowledgments}
We wish to thank S. Chattopadhyay, S. Gautam, K. V. P. Latha, B. Sahoo and
S. A. Silotri  for useful discussions. DA gratefully acknowledges 
discussions with D. Mukherjee and B. P. Das, and with D. Budker during his 
visit to Berkeley as part of the Indo-US exchange project jointly
funded by DST, India and NSF, USA. We thank I. Lindgren for valuable comments
and H. Merlitz for careful reading of the manuscript, his suggestions have
redefined the scope of the manuscript. The results presented in the paper
are based on computations using the HPC cluster at Physical Research 
Laboratory, Ahmedabad. 
\end{acknowledgments}


\end{document}